# Molecular theory of graphene oxide

**Elena F. Sheka,\* Nadezhda A. Popova**

*Russian Peoples' Friendship University, Moscow, 117198 Russia*

Applying to graphene oxides, molecular theory of graphene is based on the oxide molecular origin when it is considered as a final product in the succession of a graphene molecule polyderivatives related to a particular oxidation reaction. The graphene oxide structure is created in due course of calculations following the algorithms that take into account the graphene molecules natural radicalization, correlation of odd electrons, an extremely strong influence of structure on properties, a sharp response of the molecule behavior on small action of external factors. Taking together, the theory facilities has allowed for getting a clear, transparent and understandable explanation of hot points of the graphene oxide chemistry and suggesting reliable models of both chemically produced and chemically reduced graphene oxides.

## Problems of the graphene oxide chemistry from the computational viewpoint

One of the crucial bottlenecks for the application of graphene-based systems in materials science is their mass production. Meeting the requirements, graphene oxide (GO) has been considered widely as a prominent precursor and a starting material for the synthesis of this processable material (see recent scrupulous reviews [1-6]). On the other hand, GO turned out to resent an extremely interesting benchmark system of modified graphene and became the main object of the current graphene chemistry [4, 6]. The GO chemistry is extremely complicated and highlights all peculiarities and problems of the graphene chemistry in the largest extent.

In contrast to the current graphene science, the GO chemistry is mainly experimental. Actually, it is more than 150 years old and has collected a lot of important features of both chemical products and chemical processes. Nevertheless, in spite of the long-live studying and of high interest to the field for the last years, a lot of it has still remained unclear and ambiguous. Let us look at what we know about GOs now and which hot questions have been still not answered. Addressing the problem in such a way, we are not aimed at presenting one more review on the topic but, oppositely, basing on comprehensive reviews available, would like to concentrate the attention on some particular things that seems to be important from the computational viewpoint. To facilitate the presentation, information is grouped in four large blocks, namely: i) morphology, ii) graphene oxidation as a process in general, iii) chemical composition of GOs, and iv) current theoretical approaches. To start with the presentation, let us make clear what is implied under GO. In what follows the subject presents a product of chemical reaction of a graphene sheet with oxygen-containing oxidants that involve atomic oxygen (O), hydroxyls (OH), and carboxyls (COOH) mentioned as main participants of the GO chemistry [4, 6]. Addressing experimental evidence, will remember that in many cases graphite was a real object of the relevant study. In this case, only features characteristic for a monolayer graphene will be referred to.

*Morphology*
a. As follows from the analysis of numerous observations [7-10], the structure of graphene can be easily and remarkably disordered even by partial oxidation so that chemically produced GOs are highly amorphous.
b. Most of the recent studies on local structural characterization of GO sheets [4, 11-14] have indicated large structural disorder in the carbon skeleton due to a random arrangement of oxygen functional groups.
c. Chemically derived GO consists of isolated "molecular" sp2 domains, which are present within the carbon–oxygen sp3 matrix [15, 16].

Summarizing the data, one can state that none of regular structured GO compositions has been observed so far oppositely to one of graphene hydrides known as graphane [17-19] so that crystalline packing might not be typical for GO. Moreover, chemically derived GO seems to be of molecular nature.

*Graphene oxidation as a process in general*
a. GO is understood to be partially oxidized graphene [20].
b. The saturated at% ratio of oxygen to carbon is ~0.20-0.45 [7, 8, 21, 22].
c. When GO is heated to 11000 C, there is still about 5-10 at % oxygen left [15, 23-25].
d. Most of the oxidation of the graphene lattice proceeds in a rather random manner (see [4] and references therein).

The first two findings may indicate that there are some reasons preventing from saturation of all sp2 carbon atoms of graphene substrate by oxygen while the maximum value of the oxygen content may tell about the overall stoichiometry close to C2O. The third point highlights that the coupling of oxidants with the graphene body is energetically variable and there are special additive connections for which the coupling is particularly strong. The last point highlights that addition events under oxidation might be subordinated to a particular algorithm.

*Chemical composition of graphene oxide*
a. Despite the intensive studies on the physical and chemical properties of GO, including observations at the single-atom level, the detailed atomic structures of GO still remain elusive and a precise chemical structure has not arisen [4, 26, 27].
b. Until now, no solid evidence or feasible methods have been demonstrated to show that during the synthesis graphene can selectively be oxidized by one of the regarded oxygen-containing groups [22].
c. Extensive nuclear magnetic resonance (NMR), Raman, infrared (IR), and X-ray photoelectron (XPS) spectroscopies evidence the presence of different oxidants at graphene sheets. The types of functional groups are independent of oxygen content, however, variable stoichiometry and the local arrangement of these functionalities is unclear and is still a matter of discussion [4, 7, 9, 20]. Thus, the most common opinion attributes COOH, OH, and carbonyl C=O groups to the edge of the graphene sheet, while the basal plane is considered to be mostly covered with epoxide C-O-C and OH groups.

These observations show that GO is a product of reactions of a graphene molecule with multiple oxidants and the wide range of chemical compositions present in the reactant known as ''graphene oxide'' makes isolation and rigorous characterization of the products practically impossible [4]. Variations in the degree of oxidation caused by differences in starting materials (principally the graphite source) or oxidation protocol can cause substantial variation in the structure and properties of the material, rendering the term ''graphite oxide'' somewhat fluid, and subject to misinterpretation [4, 28].

*Theoretical and computational approaches*

The structure-property relationship is usually the main aim of any theoretical and/or computational study. However, setting such a computational problem for GO faces serious difficulties. The matter is that due to extreme variability of the GO properties mentioned above, a large set of structural models [29-34] described in details in [4] have been proposed during a long history of the graphite/graphene oxide investigation. The models are based on the analysis of GO chemical behavior as well as on studying its spectral properties by using NMR, XPS, Fourier-transformed IR (FTIR) spectra and Raman spectroscopy [4]. As fairly mentioned in [35], there are mainly two reasons why so many GO structure models have been proposed in experiment. One reason is that GO samples vary from one batch to another under different synthesis conditions. Another reason is that assignment of spectroscopic data has been based on experiences on other reference molecules and materials and thus may not be very accurate. Under these conditions, considerable expectations of theoretical studies assistance in the solution of this ambiguity of a 'fluid' GO chemistry are quite obvious. In view of this, the computation strategy of GO faced a choice of either finding supports in favor of one of the suggested models, for which spectral data provide the main pool of data for comparison with calculations, or proposing a new one.

Analyzing intense modeling, which was performed during last years, one can conclude that the realized computational strategy followed the first way. The obtained results have been recently extensively reviewed by Lu and Li [35]. Describing the computation approaches in view of the problems related to structural models and spectral data, the authors suggested grouping the available data in two parts with respect to these two topics, which well facilitates the data presentation. Briefly reviewing the data, we will follow this separation.

The first group, called as 'first-principle energetics' [35], covers structure study when analyzing energy of different structural compositions based on known structure models. In the graphene science, all the calculations of this kind [36-42] are mainly performed in the 'solid-state' approach based on the 'first-principle' closed-shell DFT approximation complemented by artificial periodic boundary conditions. In the case of GOs, the latter greatly weakens the reliability and predictability of the performed calculations, since, as shown earlier, crystalline ordering is not a characteristic of amorphous GO [34], on one hand, and GO is actually a molecule, on the other. For this and other reasons, the case of which will be considered later, the performed calculations have been unable to conclusively distinguish a preferable structure model.

The second group assembles papers attributed by the review authors to computational spectroscopy [35]. In contrast to the energetic group, it was expected that the computational spectroscopy would provide information that can be directly compared with experiment. Analogous 'first-principle' solid-state studies, based on the closed-shell DFT approach, have been performed for XPS [43], 13C NMR [44, 45], and Raman spectroscopy [46, 47]. However, as in the case of 'first-principle energetics', the obtained results were not able to distinguish the most reliable structure model among the available ones. Moreover, a clear controversy between 'first-principle energetics' and computational spectroscopy has been pointed out [35, 48]. Thus, when the former supports the wide existence of hydroxyl chains in GO basal planes GO [38, 39, 41], the latter excludes such a possibility [48]. When looking for possible reasons for the controversy among the solution and phonon contributions, as well as kinetic effect, only suggestion on kinetically constrained metastable structure of GO turned out to be valuable [48].

Therefore, the computational study of GO, based on the concept 'from a given crystal cell structure to reliable properties' has resulted in the statement about kinetically constrained metastable nature of GO. However, albeit variable but nevertheless quite reproducible when the synthesis protocol is strictly kept, the real GO chemistry does not allow for accepting this conclusion. Looking for an alternative has forced us to think about other reasons for the computational approach failure. Among such reasons, first one concerns the used DFT approach. Once called 'first-principle', it is not the case since it is highly empirically adopted due to strong dependence on the functionals in use that, in their turns, are adopted to the study by using severe empirical fitting. Second, the closed-shell version of DFT used in the studies is not applicable to the $sp^2$ graphene system due to considerable correlation of odd electrons of its carbon atoms [49, 50]. The considered closed-shell solutions do not correspond to those of the lowest energy and their energy exceed the lower one at least by ~30% that is absolutely catastrophic for the case since the 'fluid' behavior of graphene and/or GO is highly energy dependent. Third, the model structures of the studies are practically just voluntary drawn and taken for granted without any doubt concerning the possibility of their existence. Fourth, the solid-state approach has been applied to the description of molecular objects. Providing the above said, it is not surprising that this kind of theoretical approach has failed thus showing that the computational strategy of 'a voluntary given structure - properties' is not efficient for solving complicated problems of the GO chemistry.

As said above, the used DFT approach is based on the solid state microscopic theory of quasiparticles in 2D space. It is well known that, in this case, the computational results strongly depend on the unit shell under study. However, in the predominant majority of cases, even related to calculations of graphene, for which the 2D crystalline structure is out of doubts, no serious analysis has been done concerning the shape of the cell to be the most appropriate to the case.

It should be noted as well that graphene is not a simple crystal with two atoms per unit cell and hexagonal pattern of the atom distribution, but "its structure is one-atom-thick planar sheets of $sp^2$-bonded carbon atoms that are densely packed in a honeycomb crystal lattice" [51]. The evident molecular-crystal duality of graphene highlights a particular role of benzenoid building blocks, in the framework of which only we can speak about $sp^2$ carbon atoms. This aspect of the graphene science is the subject of molecular theory of graphene and graphene-based materials [49, 50]. In the framework of this theory, each GO presents a variable chemical polyderivative of a pristine $sp^2$ graphene molecule. At the same time, the theory provides a direct way of tracing the derivatives formation in due course of a stepwise addition of the corresponding addends to the pristine molecule. Computational stepwise hydrogenation, fluorination, cyanation, and aziridination of fullerene $C_{60}$ [49, 52-54] as well as hydrogenation of a graphene molecule [18], convincingly approved by comparison with experiment, are attractive examples of computational synthesis of polyderivatives of sp² nanocarbons. The other strong point of molecular theory of graphene is its ability to exhibit and take into account a topological character of chemical reactions involving graphene [55-57]. Therefore there are convincing grounds for believing that similar approach to the problem of GO will be successful as well.

In view of the molecular theory, the hot points of the GO chemistry can be presented as the following questions:
- What is the role of individual oxidants, such as O, OH, and COOH, in the oxidation of a graphene molecule?
- How does the coupling of the oxidants with the molecule depend on the local place of the addition?
- How is the highest degree of the derivatization under graphene molecule oxidation?
- What is the structure of the carbon skeleton of GO and can be the regular structure achieved?
- What chemical composition of GO can be obtained under conditions close to experimental ones?

Answering each of the questions presents a topic for a valuable computational study. A temptation to more or less answer all the questions simultaneously raises a problem of a system approach, or, by other words, of an extended computational experiment. The computational molecular theory of graphene is the very feasible method in the framework of which the approach can be realized. The latter means that not selected individual computations but a solution of a number of particularly arranged computational problems is the aim of the computational experiment. Evidently, the bigger set, the more colors participate in drawing the image of the GO chemistry. The current paper presents the first attempt of such system computational approach. To obtain a rather complete picture, about 400 computational jobs had to be performed. The obtained results allow explanation of the majority of unclear experimental findings and highlight the efficacy of a new computational strategy of GO named "molecular derivative structure - properties".

**General comments on oxidation**

The most particular feature of the graphene oxidation concerns a multi-oxidant character of this chemical reaction. Actually, O, OH, COOH, and partially H are the main addends that may attack the graphene molecule [4]. Obviously, water molecules play a very important role in the GO production as well [2, 4, 6, 58]. However, the latter do not form chemical bonds with graphene [18] so one may not add the molecule to the above list of reactants.

From general regularities of intermolecular interaction, which is responsible for the GO formation, the process might be characterized by two energetic parameters, namely: the coupling energy $E_{cpl}(GO)$, which provides the energy profitability of the final product formation, and the barrier energy $E_{barr}(GO)$ that points to the extent of difficulties for the reaction kinetics. The coupling energy can be determined as

$$E_{cpl}(GO) = \Delta H(GO) - \Delta H(Gr) - \Delta H(Oxd) \tag{1}$$

Here, $\Delta H(GO)$, $\Delta H(Gr)$, and $\Delta H(Oxd)$ are heats of formation of GO, pristine graphene molecule, and one of the above oxidants, respectively. All the three quantities as well as $E_{cpl}(GO)$ correspond to intermolecular distances between the molecule and oxidants those provide the formation of chemically bound product. $\Delta H(Oxd)$ forms a series of +59.559, +0.633, -54.704, and +52.234 kcal/mol for O, OH, COOH, and H, respectively. One should draw attention to the big negative value of $\Delta H(Oxd)$ for COOH in contrast to positive values for other species. Obviously, this point does not favor the carboxyl group from the very beginning and makes its attachment to graphene molecule the least evident.

The energy $E_{barr}(GO)$ is a characteristic of the barrier profile that describes the dependence of the intermolecular interaction on the intermolecular distance and that should be determined in each case individually similarly as was done, say, in the case of attaching C60 molecule to a set of nanocarbons [55]. In the case of GO, it was shown [59] that the attachment of either O or OH to the graphene molecule occurs without barrier while COOH and H can be attached to the molecule when overcoming barriers of 1.7 and 3.1 eV (attachments to one of zigzag edge atoms in both cases), respectively. Evidently, this finding highlights an obvious preference for O and OH to be attached to graphene. In contrast, the barrier availability makes COOH and H additions less probable. It seems that these very features of the energy characteristics of COOH provide its low, if any, presence in real GO products [4]. Actually, all the chemists suppose the absence of the oxidant among those attached to basal plane of graphene [4, 6]. A large part of investigators accepts a similar situation to be held at the graphene edges as well while the others suppose the presence of the oxidant in this region. The matter concerns an ambiguity in the interpretation of the FTIR absorption spectra of the studied GO due to problematic attribution of particular bands to characteristic C-O and C=O vibrations [4, 60]. To convert the presented speculations into a realistic vision of the GO chemistry, it is necessary to disclose the oxidation at a microscopic level. Results, presented in next Sections, will demonstrate how it can be realized in the framework of molecular theory of graphene.

**Molecular theory grounds for a system computational experiment**

In view of a significant correlation of odd electron in graphene [50], peculiarities of the graphene chemistry can be exhibited at the quantitative level, much as this has been done for fullerenes [49]. As was shown by Takatsuka, Fueno, and Yamaguchi [61], the correlation of weakly interacting electrons is manifested through a density matrix, named as the distribution of 'odd' electrons,

$$D(r|r') = 2\rho(r|r') - \int \rho(r|r'')\rho(r''|r')dr''. \tag{2}$$

The function $D(r|r')$ was proven to be a suitable tool to describe the spatial separation of electrons with opposite spins and its trace

$$N_D = tr D(r|r') \tag{3}$$

was interpreted as the total number of these electrons [62, 63]. The authors suggested $N_D$ to manifest the radical character of the species under investigation. Over twenty years later, Staroverov and Davidson changed the term by the 'distribution of effectively unpaired electrons' [63, 64] emphasizing that not all odd electrons may be taken off the covalent bonding. Even Takatsuka et al. mentioned [61] that the function $D(r|r')$ could be subjected to the population analysis within the framework of the Mulliken partitioning scheme. In the case of a single Slater determinant, Eq. 3 takes the form [62]

$$N_D = tr DS, \tag{4}$$

where

$$DS = 2PS - (PS)^2. \tag{5}$$

Here, $D$ is the spin density matrix $D = P^\alpha - P^\beta$ while $P = P^\alpha + P^\beta$ is a standard density matrix in the atomic orbital basis, and $S$ is the orbital overlap matrix ($\alpha$ and $\beta$ mark different spins). The population of effectively unpaired electrons on atom A is obtained by partitioning the diagonal of the matrix $DS$ as

$$D_A = \sum_{\mu \in A} (DS)_{\mu\mu}, \tag{6}$$

so that

$$N_D = \sum_A D_A. \tag{7}$$

Staroverov and Davidson showed [63] that the atomic population $D_A$ was close to the Mayer free valence index in general case, while in the singlet state $D_A$ and $F_A$ are identical. Thus, a plot of $D_A$ over atoms gives a visual picture of the actual radical electrons distribution [63], which, in its turn, exhibits atoms with enhanced chemical reactivity.

A correct consideration of the odd electron correlation requires multi-determinant computational schemes. However, a low efficacy of the modern computational programs of this kind greatly limits their practical application. On the other hand, properly arranged and particularly adopted single-determinant approximations may be of great advantage. The unrestricted broken symmetry (UBS) approximation, suggested by Noodleman [65], belongs to such a type. Realized in the framework of the Hartree-Fock computational scheme (UBS HF), it offers an efficient computational technique that is able to highlight a lot of peculiarities of the odd electron behavior.

The effectively unpaired electron population is definitely connected with the spin contamination of the UBS solution. In the case of UBS HF (UHF) scheme, there is a straight relation between $N_D$ and squared spin $\langle \hat{S}^2 \rangle$ [63]

$$N_D = 2\left( \langle \hat{S}^2 \rangle - \frac{(N^\alpha - N^\beta)^2}{4} \right), \tag{8}$$

where,

$$\langle \hat{S}^2 \rangle = \left( \frac{(N^\alpha - N^\beta)^2}{4} \right) + \frac{N^\alpha + N^\beta}{2} - \sum_i^{N^\alpha} \sum_j^{N^\beta} |\langle \phi_i | \phi_j \rangle|^2. \tag{9}$$

Here, $\phi_i$ and $\phi_j$ are atomic orbitals; $N^\alpha$ and $N^\beta$ are the numbers of electrons with spin $\alpha$ and $\beta$, respectively.

If UHF computations are realized in the *NDDO* approximation (the basis for AM1/PM3 semiempirical techniques) [66], a zero overlap of orbitals leads to $S = I$ in Eq. 5, where I is the identity matrix. The spin density matrix $D$ assumes the form

$$D = (P^\alpha - P^\beta)^2. \tag{10}$$

The elements of the density matrices $P_{ij}^{\alpha(\beta)}$ can be written in terms of eigenvectors of the UHF solution

$$P_{ij}^{\alpha(\beta)} = \sum_k^{N^{\alpha(\beta)}} C_{ik}^{\alpha(\beta)} * C_{jk}^{\alpha(\beta)}. \tag{11}$$

Expression for $\langle \hat{S}^2 \rangle$ has the form [67]

$$\langle \hat{S}^2 \rangle = \left( \frac{(N^\alpha - N^\beta)^2}{4} \right) + \frac{N^\alpha + N^\beta}{2} - \sum_{i,j=1}^{NORBS} P_{ij}^\alpha P_{ij}^\beta \ . \tag{12}$$

Within the framework of the *NDDO* approach, the HF-based total $N_D$ and atomic $N_{DA}$ populations of effectively unpaired electrons take the form [68]

$$N_D = \sum_A N_{DA} = \sum_{i,j=1}^{NORBS} D_{ij} \tag{13}$$

and

$$N_{DA} = \sum_{i \in A} \sum_{B=1}^{NAT} \sum_{j \in B} D_{ij} \ . \tag{14}$$

Here, $D_{ij}$ are elements of spin density matrix $D$ that presents a measure of the electron correlation [61, 63, 69].

Explicit expressions (13) and (14) are the consequence of the UHF wave-function-based character. Since the corresponding coordinate wave functions are subordinated to definite permutation symmetry, each value of spin $S$ corresponds to a definite expectation value of energy [70]. Oppositely, the electron density is invariant to the permutation symmetry. The latter causes a serious spin problem for the UBS DFT (UDFT) [70, 71]. Additionally, the spin density $D(r|r')$ of the UDFT depends on spin-dependent exchange and correlation functionals and can be expressed analytically in the former case only [70]. Since the exchange-correlation composition deviates from one method to the other, the spin density is not fixed and deviates alongside with the composition. Serious UDFT problems are known as well in the relevance to $\langle \hat{S} \rangle^2$ calculations [72, 73]. These obvious shortcomings make the UDFT approach practically inapplicable in the case when the correlation of weakly interacting electrons is significant. Certain optimism is connected with a particular view on the structure of the density matrix of effectively unpaired electrons developed by the Spanish-Argentine group [69, 74,75] from one hand and new facilities offered by Yamagouchi's approximately spin-projected geometry optimization method intensely developed by a Japanese team [76, 77], from the other. By sure, this will give a possibility to describe the electron correlation at the density theory level more thoroughly.

In the singlet state, the $N_{DA}$ values are identical to the atom free valences [63] and thus exhibit the atomic chemical susceptibility (ACS) [78, 79]. The $N_{DA}$ distribution over atoms plots a 'chemical portrait' of the studied molecule, whose analysis allows for making a definite choice of the target atom with the highest $N_{DA}$ value to be subordinated to chemical attack by an external addend. This circumstance is the main consequence of the odd electron correlation in graphene in regards its chemical modification. Ignoring the correlation has resulted in a common conclusion about chemical inertness of the graphene atoms with the only exclusion concerning edge atoms. Owing to this, a practicing computationist does not know the place of both the first and consequent chemical attacks to be possible on, say, the basal plane and has to perform a large number of calculations sorting them out over the atoms by using the lowest-total-energy criterion (see, for example, [80]). In contrast, basing of the $N_{DA}$ value as a quantitative pointer of the target atom at any step of the chemical attack, one can suggest the algorithmic 'computational syntheses' of the molecule derivatives [52] that shows its efficiency for the computational chemistry of fullerene $C_{60}$ [49] and graphene hydrogenation [18].

In what follows the algorithm-in-action will be illustrated by the example of the oxidation of a graphene molecule (5, 5) NGr. The molecule was previously used when studying graphene hydrogenation [18] and mechanochemical deformation under uniaxial tension [81, 82], which has greatly contributed into understanding general peculiarities of its chemical behavior. All the following computations have been performed in the broken symmetry approach by using unrestricted Hartree-Fock computational scheme implemented in CLUSTER-Z1 codes based on semiempirical AM1 approach (a detailed description of the strategy of the computational consideration of sp2 nanocarbons has been summarized in [50]). Application of this effective program has allowed for performing ~ 400 computational jobs during a two-month session for a number of GO molecules differing by atom number from 66 to 218. Conclusions presented below are based on a comparable study of the results obtained.

## Algorithmic computational design of graphene polyoxides

The equilibrium structure of the pristine graphene molecule (5,5) NGr is shown in Fig.1a. The molecule is a free standing rectangular fragment (membrane in terms of [18]) of a graphene sheet containing 5 and 5 benzenoid units along armchair and zigzag edges. Panel b in the figure exhibits the molecule ACS image map that presents the distribution of atomically-matched effectively unpaired electrons $N_{DA}$ over the molecule atoms attributed to the atoms positions, thus visualizing the 'chemical portrait' of the molecule. Different ACS values are plotted in different coloring according to the attached scale. The absolute ACS values are shown in panel c according to the atom numbering in the output file. As seen in the figure, 22 edge atoms involving 2x5 zg and 2x6 ach ones have the highest ACS thus marking the perimeter as the most active chemical space of the molecule. These atoms are highlighted by the fact that each of them posses two odd electrons, the interaction between which is obviously weaker than that for the basal atoms. Providing the latter, the electron correlation and the extent of the electron unpairing are the highest for these atoms, besides bigger for zg edges than for ach ones. These general features of the molecule chemical portrait are characteristic for any graphene fragment of different size and shape.

Stepwise addition of each of the above listed oxidants to the molecule provides controlling wished reactions. For sp$^2$ nanocarbons, these reactions are governed by a particular computational algorithm [49, 52]. A spatial distribution of effectively unpaired electrons $N_{DA}$ over the carbon skeleton of the molecule lays the algorithm foundation. The atomically mapped high rank $N_{DA}$ values are taken as pointers of target atoms at each reaction step. According to Fig. 1c, all the reactions for the (5,5) NGr molecule will start on zigzag atom 14 (star-marked in Fig.1c) and their further developing will depend on the redistribution of the $N_{DA}$ values over the remaining carbon atoms after the first addition of either O, OH, and COOH (H is taken for comparison) to this atom is completed. The redistribution is caused by changing C-C bond lengths of the molecule skeleton. After the first addition, the $N_{DA}$ maps of the first derivatives of the molecule, related to the series of the above addends, are headed by atoms 59, 51, 55, and 55. All the atoms are located at the opposite zigzag edge of the molecule. However, in this series of target atoms for the second step, the starting points coincide only for OH and COOH while atomic addition of hydrogen and oxygen occurs at other places. This finding is very important since the difference in the target atoms number highlights the difference in the odd electron cloud response on different chemical additions to the molecule thus exhibiting a particular feature of a conventional 'chemical modification' of graphene.

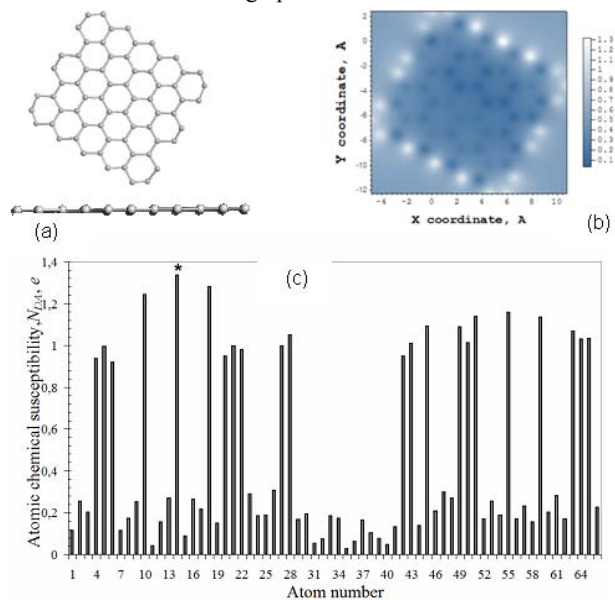

**Fig. 1**. Top and side views of the equilibrium structure of the (5,5) NGr molecule (a); ACS image map (b); ACS distribution over atoms according to their numbering in the output file (c) [18].

The molecule (5, 5) NGr has 88 odd electrons, 44 of which belong to 22 edge atoms and the remainder 44 ones to 44 basal atoms. Part of these electrons are involved in the covalent interatomic bonding, while 31.038 of them, which constitute the total number of effectively unpaired electrons $N_D$ [50], provide a radical nature of the molecule and determines the molecule chemical susceptibility. Obviously, $N_D$ determines the molecule reactivity pool that once distributed over the molecule atoms provides the formation of polyderivatives, while, on the other hand, is gradually working out in due course of the derivatives formation. The number of steps needed for $N_D$ to be completely worked out, determines the extent of the polyderivatization to be achieved. The stepwise fluorination and hydrogenation of fullerene $C_{60}$ have shown [52, 53] that the relevant addition reactions stop when $N_D$ becomes zero. In computational practice, the step number depends on both the valence state of addends and the manner of their addition. Thus, in the case of (5, 5) NGr, 44 steps were needed to provide the stepwise addition of hydrogen atoms to the molecule edge atoms [18]. Similarly, 42 steps were needed to saturate chemical reactivity of the basal-plane carbon atoms by this addend. In the case of a homo-oxidant addition, a similar total number of steps might be needed to complete the molecule oxidation by either OH or COOH. For atomic oxygen the number of steps should be twice less. Obviously, hetero-oxidant additions will require the step number between the two limit cases.

The performed experiment has been arranged for the maximum number of possible polyderivative configurations to be considered. A comparable study of the results related to different configurations of GOs has led the foundation for the suggested conclusions. To facilitate the description of the results, we divide the data into two sets. The first set, which corresponds to the first stage of the oxidation in practice, covers first 22 steps that involve edge atoms of the molecule only, on one hand, and all the oxidants, on the other. The second step combines results related to both edge and basal atoms for oxidation by OH and COOH and basal atoms only when attaching atomic oxygen. We shall refer to these data sets as to the first-stage and second-stage oxidation of the (5, 5) NGr molecule.

## The first-stage oxidation of the (5, 5) NGr molecule

**Homo-oxidant action**

Figure 2 presents equilibrium structures of the O-, OH-, and COOH- multi-fold GOs at the first five and the 22nd steps of oxidation. For comparison, a similar first stage of the (5, 5) NGr molecule hydrogenation is added. A complete hydrogenation of the molecule is presented in [18]. In all the cases, each next step of a subsequent addition was governed by the highest rank NDA value at a preceding step. As seen in the figure, the difference in the starting points at the second step has actually led to the different developing of the hydrogenation and oxidation processes in space.

In the case of O-GOs, the obtained 22-fold GO (GO I) presents a completed framing of the molecule by these oxidant. In the case of OH-GOs, a successive OH- framing continues up to the 12th step. According to the ACS map, the 13th step starts at atom 50 (see star marked atom in Fig.3a). After optimization of thus formed 13-fold OH-GO, its equilibrium structure reveals the dissociation of one previously attached OH group and the formation of C-H and C=O bonds instead (see Fig.3b). All the next steps of oxidation up to the 21st one do not cause any similar transformation and the final GO II shown in the last row in Fig.2 was obtained. If one tries 'to correct' the situation hold at the 13th step by hand and restores the dissociated hydroxyl at previous place, optimization of the structure results in splitting one of the first attached hydroxyls from the carbon core of the molecule at all (see Fig. 3c). The hydroxyl dissociation and splitting are caused by a considerable deformation of the skeleton shape. The molecule becomes non-planar practically since the third step. By the 13th step, the molecule plane curvature is so big that interatomic distances change drastically, particularly, in the area of edge atoms. Due to the dependence of odd correlation from the distance [83], the strength of the chemical interaction for these atoms strongly varies, which causes both dissociation and splitting OH groups. The skeleton deformation becomes more smoothed by the end of framing. That is why, substitution of oxygen and hydrogen addends in the final 21-fold structure of GO II in Fig.2 by two hydroxyls results in a comfortable 22-fold OH-framed GO IV shown in Fig.3d. In the case of COOH-oxides, sterical constrains additionally complicate framing so that the final structure of the 22-fold COOH-framed GO III shown at the bottom of Fig.2 does not seem strange.

As might be expected, the first-stage framing considerably suppresses the chemical activity of the molecule edge atoms, but, evidently, not to the end. Shown in Fig.4 highlights changes in the ACS maps of all the four derivatives shown at the bottom of Fig.2. The changing can be traced by viewing both image maps of the molecule and ACS/Step number plottings. The latter are drawn over that one of the pristine molecule so that one can see ACS changing caused by the 22-fold substitution more vividly. The numbers of targeting carbon atoms are 4, 5, 6, 10, 14, 18, 20, 21, 22, 27, 28, 42, 43, 45, 49, 50, 51, 55, 59, 63, 64, and 65.

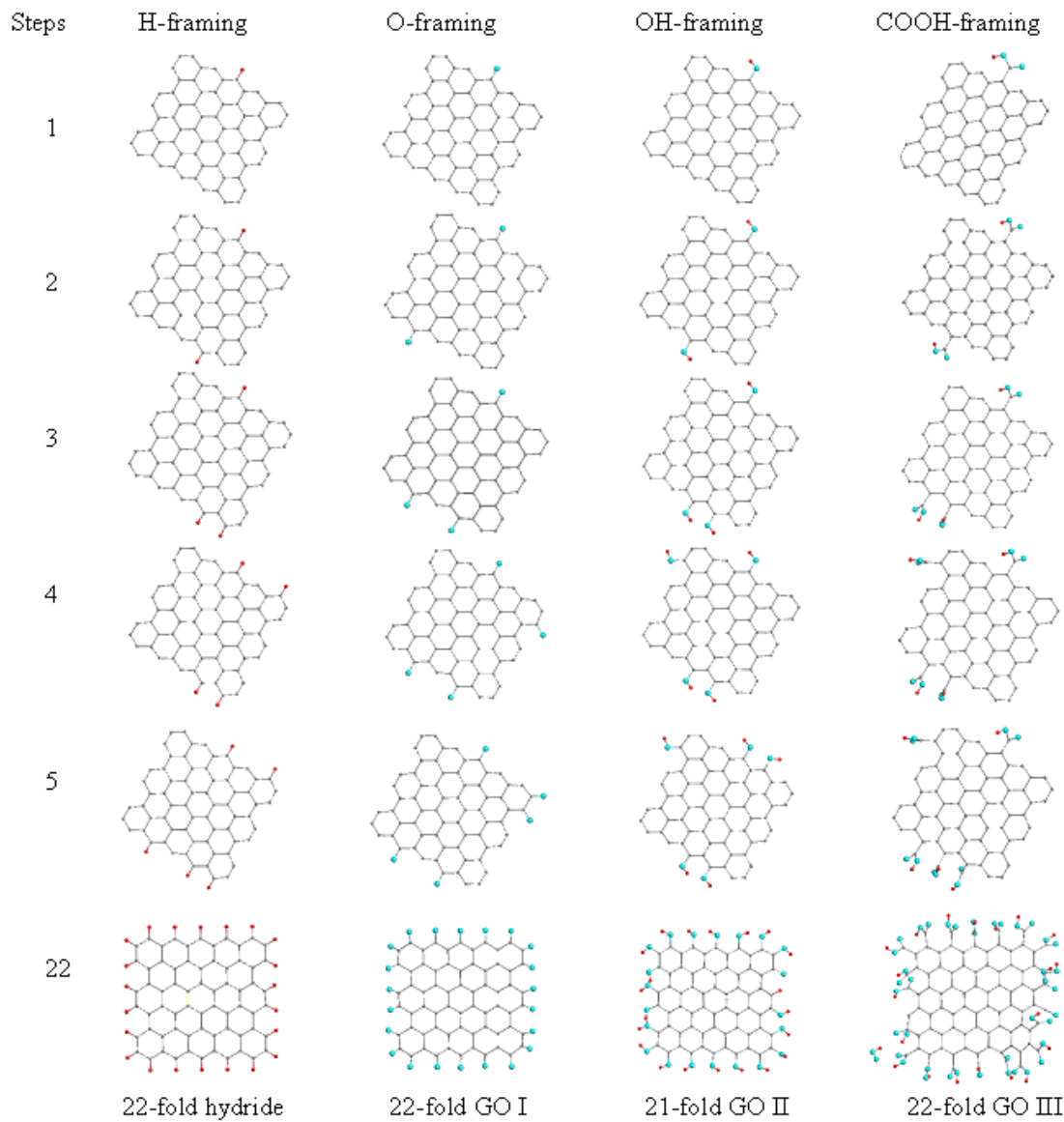

**Fig. 2**. Equilibrium structures of (5,5) NGr polyhydrides and polyoxides related to the 1st, 2nd, 3rd, 4th, 5th, and 22nd (21st in the case of OH-framing, see text) steps of first-stage reactions. Roman figures mark the obtained GOs.

As seen in Fig.4a, the 22-fold H-framing results in ~4-fold decreasing ACS of the edge atoms making them comparable by the activity with basal atoms. Simultaneously, the framing causes a redistribution of ACS over basal atoms. Therefore, the continuation of the molecule hydrogenation must concern both edge and basal atoms practically at equal level. However, as was shown [18], the next 22 steps of hydrogenation concerned edge atoms only resulting in a complete suppressing their chemical activity for the 44-fold graphene hydride. Afterwards, the hydrogenation moved to the basal plane.

In the case of O-GOs, carbonyls, formed over the molecule perimeter, provide practically full suppression of the chemical activity of the edge atoms. Only edge atom 43 shown by bright spot at the image map of the GO goes out of the rule due to exclusively large length of the relevant C=O bond. The bond elongation is caused by the optimization of all the oxygen atoms accommodation at the perimeter of the (5, 5) NGr molecule under the condition of the least perturbation of the carbon skeleton structure. The further oxidation of the molecule, starting at the 23rd step, occurs at the basal plane.

The distribution of the chemical activity of the 21-fold OH-GO in Fig. 4c looks like that one of the molecule hydride. The activity of the edge atoms is greatly suppressed and made practically equal to the activity of basal atoms. The activity of only one edge atom that forms carbonyl group is practically zero (see black spot at the image map in Fig. 4c). Both edge and basal atoms should be taken into account when considering a further oxidation.

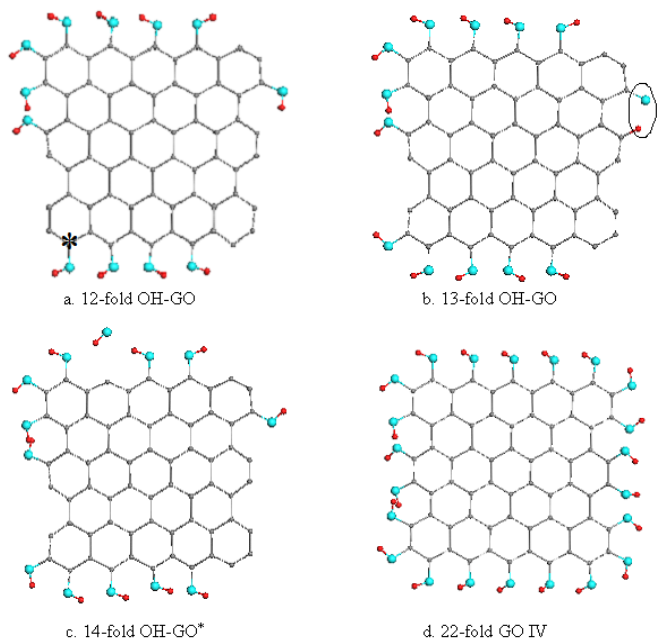

a. 12-fold OH-GO  
b. 13-fold OH-GO  
c. 14-fold OH-GO*  
d. 22-fold GO IV

**Fig. 3**. Equilibrium structures of multi-fold OH-framed GOs (see text).

The succession of the molecule framing by 22 COOH units is quite similar to the discussed above with the only difference concerning atoms 42 and 43 that remain non-attached (two bright spots in the image map in Fig.4d) due to sterical constrains for the COOH oxidant. A comparison of the image ACS maps presented in Fig. 4a, c, and d shows how differently influence the added addends on the electronic properties of final products in spite of their mono-valence ability in all the cases.

The analysis of homo-oxidant framing of the (5, 5) NGr molecule is not complete without considering the energetic regularities that accompany the relevant GOs formation. Figure 5 presents a set of the dependences of perstep coupling energies (PCEs) on the step number. The PSE is determined as

$$E_{cpl}^{pst}(GO_n) = \Delta H(GO_n) - \Delta H(GO_{n-1}) - \Delta H(Oxd) \cdot \quad (15)$$

Here, $\Delta H(GO_n)$ and $\Delta H(GO_{n-1})$ are heats of formation of considered GOs at the $n^{th}$ and $(n-1)^{th}$ steps of oxidation, respectively. PCEs are much more appropriate for a comparable study of different GOs than total energies since the latter contain a number of contributions that make the GO total energies too cumbersome.

Analyzing data presented in Fig.5, one can conclude the following.

1. The dependences of $E_{cpl}^{pst}(GO_n)$ on the step number have much in common for all the oxidants. Thus, they show non-regular oscillating behavior that is characteristic for the same stage of hydrogenation as well once provided supposedly by a particular topology of the (5, 5) NGr molecule that reflects a changeable disturbance of the molecule carbon skeleton in due course of oxidation.

2. The molecule framing by carbonyls is evidently the most energetically favorable while, oppositely, COOHs provide the least stable configurations. The coupling characteristic for OH-GOs is in between these two limit cases and only at the 13th step of the OH-GO is the largest one among the others due to dissociation of the hydroxyl group discussed earlier.

3. The difference in PSE between COOH-GOs and OH- and O-GOs is not too drastic as might be expected on the basis of the discussion presented in Section 2, particularly, at first steps of the oxidation.

According to this finding, one should expect that carbonyls play the dominant role in the framing of GO molecules in practice. In contrast, the most spread opinion supports the scheme of the GO chemical composition suggested by Lerf and Klinovsky [33, 84] where not carbonyls but epoxy groups frame the GO platelets. To check this suggestion, we have performed a few calculations looking at epoxy framing of the studied molecule. The equilibrium structures of the epoxy-GOs corresponding to the first three steps of oxidation are shown in Fig.6. Curve 4 in Fig.5 plots the related PSEs. As seen in the figure, the epoxy-GOs are not only the least stable by energy but even become energetically non-profitable since PSE becomes positive from the third step. This finding makes it possible to exclude epoxy groups from possible candidates of GO platelets framing.

**Hetero-oxidant action**

In practice, the graphene oxidation occurs in the multi-oxidant media so that the final product might present hetero-oxidant GOs. Since, as shown in the previous section, attaching different oxidants to the same edge atom of the molecule causes different continuation of the reaction; such GOs can not be obtained on the basis of a simple superposition of the data discussed before so that particularly arranged calculations are needed. Evidently, a simultaneous addition of different oxidants to the molecule edges is quite ambiguous. Seemingly, basing on the ACS-governed algorithm of the target atom selection, we could have chosen, say, three the highest rank target atoms following the indication of the NDA values list for attaching three different oxidants. However, we know by certain that any individual attachment to the carbon skeleton drastically changes the NDA listing, which makes the choice of three atoms for a simultaneous attachment quite uncertain. Simultaneous multiattachment does not make it possible as well to analyze the competition between the oxidants and to reveal the most favorable among them. Providing the facts, we consider a subsequent addition of different oxidants more suitable for the problem solution. The computations have been performed following such a procedure. Basing of the ACS-governed algorithm, we determine the first-attack target atom and successively attach to it either O, or OH and COOH oxidant. Following the largest PCE criterion, we select the only GO that meets the requirement and determine the number of the second-step attack atom by its ASC map. Afterwards, the procedure of the successive addition of the three oxidants is repeated and new GO with the largest PSE is chosen. Its ACS map shows the target atom for the next attack and so forth. Figure 7d exhibits the PSE plottings versus step number for the three oxidants. Comparing the data with those related to a homo-oxidant action shown in Fig. 5 one can conclude the following.

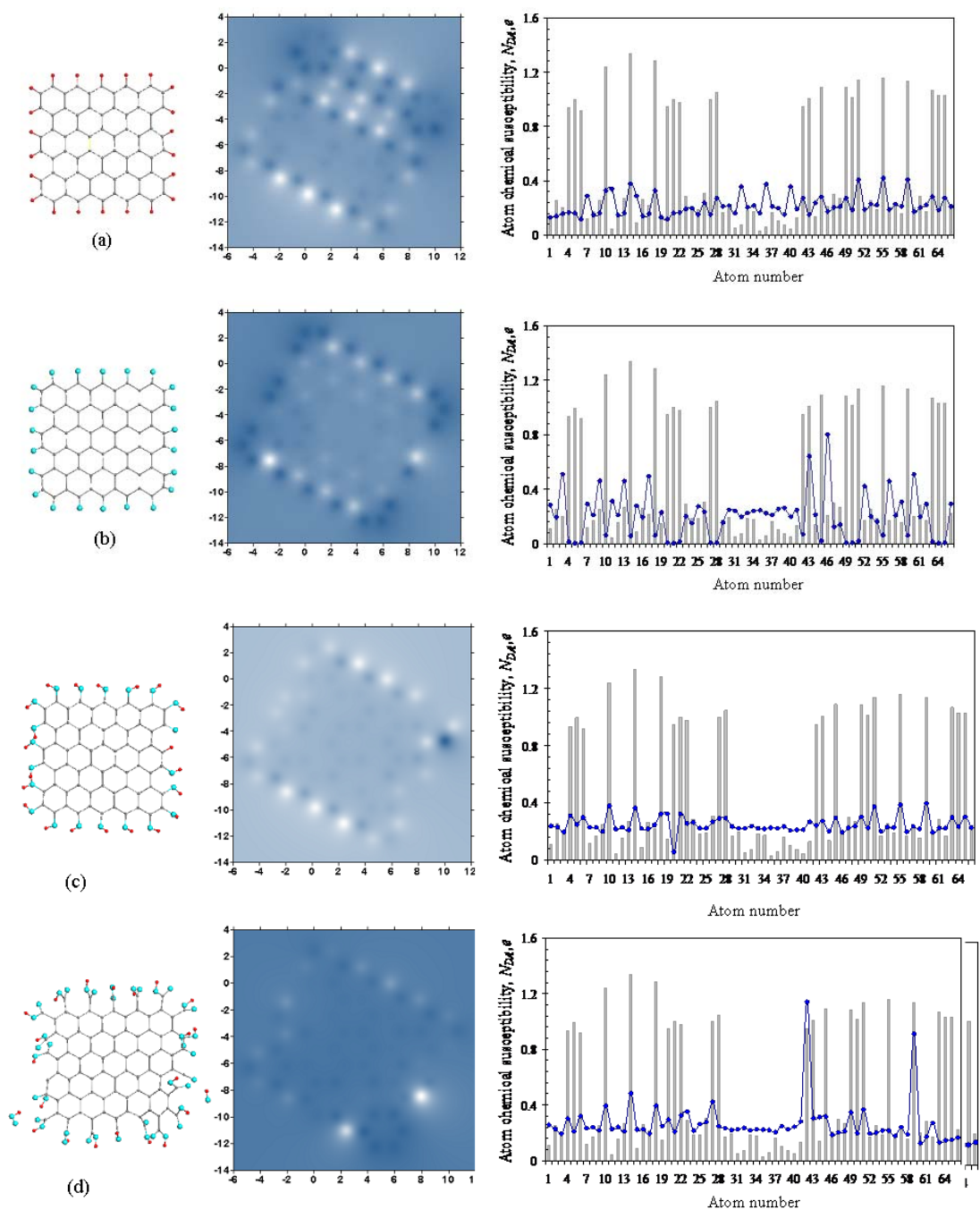

**Fig.4**. Equilibrium structures (left), image ACS maps (center) and ACS distribution versus atom number (right) of polyderivatives of the (5,5) NGr molecule: a. 22-fold hydride; b. 22-fold O-framed GO I, c. 21-fold OH-framed GO II; d. 22-fold COOH-framed GO III.

    1. Plottings related to each of the considered oxidants are quite different in the two figures. This is a consequence of the difference of the chemical composition at the edge atoms during homo- and hetero-oxidant action.
    2. Hetero-oxidant computational experiment exhibits quite undoubtedly that COOHs are not favorable for the (5, 5) NGr molecule framing.
    3. The molecule framing is predominantly provided by carbonyls and partially by hydroxyls so that the final framing composition is double-oxidant corresponding to the 22-fold GO V whose equilibrium structure is presented in Fig.7a.
    4. The carbonyl:hydroxyl ratio of 20:2 is evidently non-axiomatic and supposedly quite strongly depends on the pristine graphene molecule size. However, the predominance of carbonyls at the periphery of GO molecules should be expected in practice while the presence of COOHs is strongly doubtful.

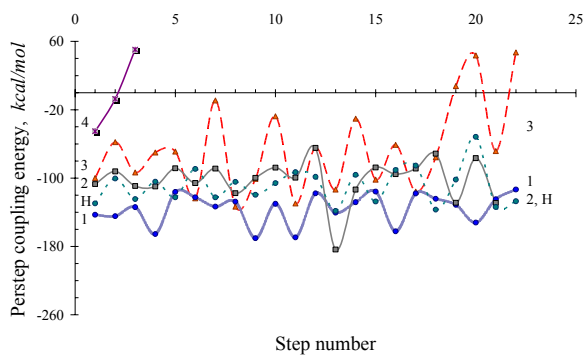

**Fig.5**. Perstep coupling energy versus step number for families of GOs obtained in due course of the first- stage oxidation. 1. GO I; 2. GO II; 3. GO III; 4. Epoxy-framed GOs (see text); H. Hydrides.

### The second-stage oxidation of the (5, 5) NGr molecule

The data presented in the previous Section, even once limited to the first steps of oxidation, convincingly show how complicated is the chemical modification of graphene in this case. Surely, the process does not stop at this level and continues further involving both edge atoms, if they are still not valence compensated, and basal ones thus becoming more and more complicated. To demonstrate the extent of complication, we have performed a few experiments based on GO I, GO II, GO IV, and GO V discussed in the previous Section. They form two sets combining GOs II and IV with OH-framing and GOs I and V with O-framing.

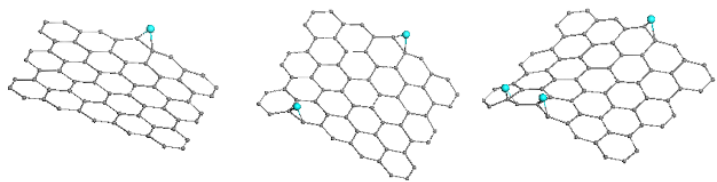

**Fig.6**. Equilibrium structures of epoxy-framed GOs at first three steps of oxidation.

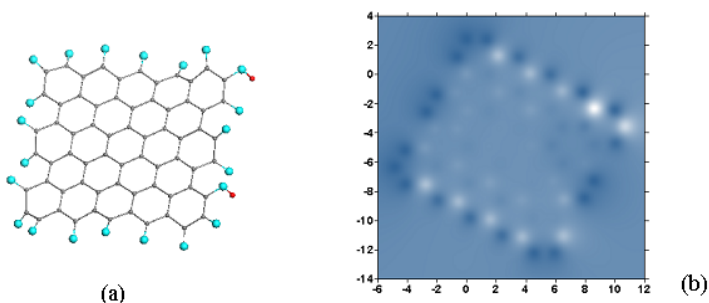

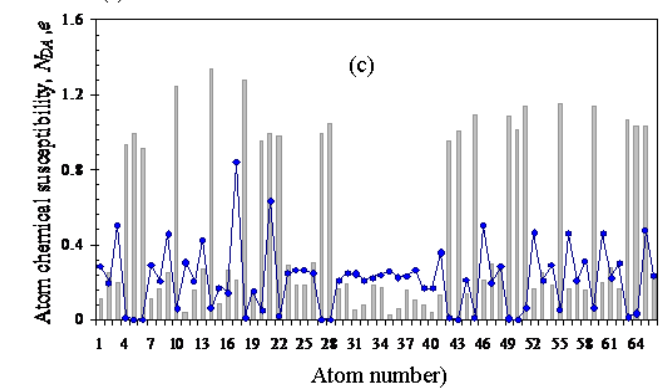

**Fig.7**. First-stage hetero-oxidant framing of the (5,5) NGr molecule. a. Equilibrium structure of 22-fold (O+OH)-framed GO V; b. ACS image map of GO V; c. ACS distribution over GO V atoms. d. Perstep coupling energy versus step number for the GO V family. 1.O-additions; 2. OH-additions; 3. COOH-additions.

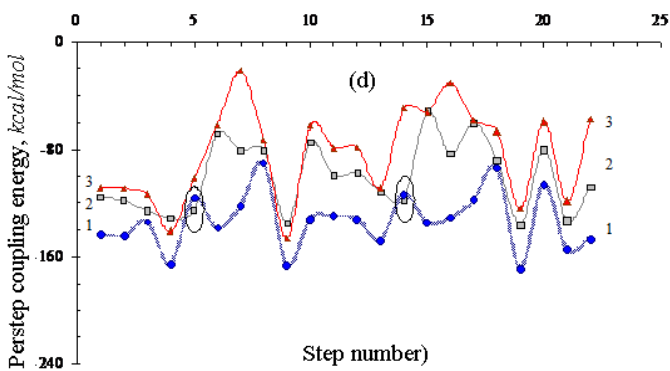

### Oxidation of hydroxyl-framed GOs

Figure 8 summarizes structural results typical for the second-stage homo-oxidant reaction (SSHMR). Image maps of GOs II and IV shown in Figs. 8a and d evidently highlight the difference in starting conditions for the oxidation prolongation. Six brightest spots on the maps distinguish six atoms situated at the zigzag edges, whose chemical activity is the most non-compensated after adjoining solitary OH groups. Obviously, the first step of the second-stage reaction will concern one of these atoms in both cases. However, the atom numbers are different which is indicated by circling. The first experiment steps concern the addition of the second OH to a sequence of edge atoms, each of which was indicated by the highest ACS in the relevant ACS map at the preceding step. However, the reaction becomes non-controlled at the fifth step. After reaching equilibrium structure at the fourth step shown in Fig.8b, the ACS map points to the prolongation of the reaction by addition of the second hydroxyl to edge atom 22 shown by circle. The followed structure optimization has revealed a destructive structure presented in Fig.8c; three hydroxyls and one hydrogen molecule have left the GO molecule while locating in the near neighborhood, a slightly stretched oxygen molecule is chemically attached to atom 59 (circled). Due to radical character, the three hydroxyls take the first three places in the ACS list thus masking the ACS field of the molecule and highly complicating the algorithmic continuation of the oxidation reaction. Additional complexity concerns a correct determination of the PSE. Therefore, we have decided to stop the reaction at this level just keeping the experiment as an illustration of complications that can be expected during graphene oxidation.

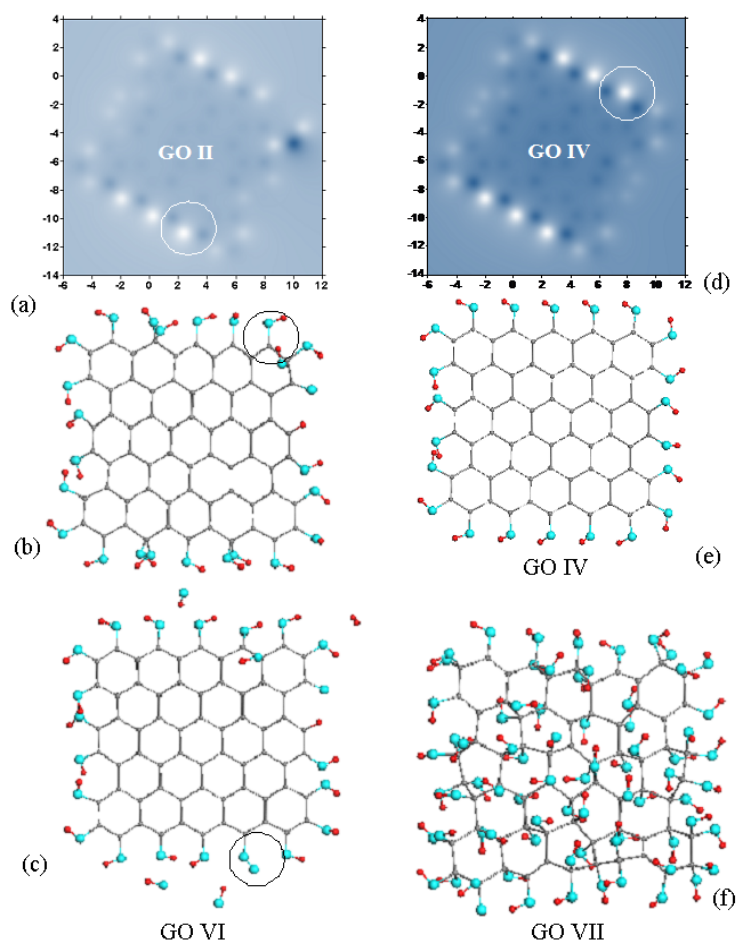

**Fig.8.** SSHMR of the OH-oxidation of the (5,5) NGr molecule. a. Image ACS map of pristine GO II. b. Equilibrium structure of the 26-fold OH-GO after the 4th step of SSHMR. c. Structure distortion of GO VI at the 5th step of SSHMR. d. Image ACS map and e. equilibrium structure of pristine GO IV. f. Equilibrium structure of GO VII after 52 steps of SSHMR.

To explain so drastic changing in structure it is necessary to accept that in spite of local attachment of the oxidant to the molecule atom 22, the final results are provided by the total system of unpaired electrons. Any new addition changes space structure due to changing contributions of sp2 and sp3 configured carbon atoms. The structure changing causes a redistribution of C-C bond lengths, which, in its turn, results in changing ACS distribution that, finally, may drastically reconfigure the coupling strength field, which is really observed. An extreme sensitivity of structure-response relationships is one of particular characteristics of the graphene electron system. Obviously, the deviation in the first stage OH framing of the molecule due to the presence of single C-H and carbonyl groups among homo-oxidant OH necklace has lent an additional sharpness to the feature.

Graphene oxide IV (see Fig.8e) was led in the foundation of the second SSHMR experiment. As previously, the first five steps were limited by the second attachments of hydroxyls to a sequence of edge atoms. In contrast to the previous case, the homogeneous first stage framing and changing the place of the first attack at the second stage assisted in avoiding disruptive events occurred with GO VI so that none of the oxidants or some parts of them has left the molecule and a controlled oxidation reaction has proceeded up to the 52nd step. At the sixth step, the reaction moved into basal plane for one step, then came back to edge atoms for two steps, then returned to basal plane (2 steps), to edge atoms (2 steps) and was located in the basal plane for the next fifteen steps. Afterwards, during succeeding twenty five steps the reaction seven times came back to edge atoms, mainly staying in basal plane. The 74-fold GO VII shown in Fig.8f corresponds to equilibrium structure after termination of the 52nd step of SSHMR. All additions occurred at basal plane were considered for two position of the hydroxyls, namely, above and under the plane and the choice of the best configuration was subordinated to the lowest energy criterion.

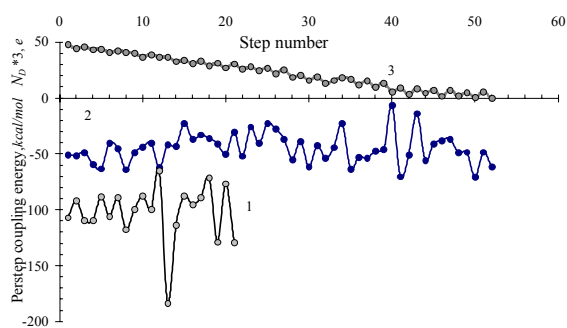

**Fig. 9.** Perstep coupling energy versus step number for GO VII family in due course of the first- stage (1) and second-stage (2) oxidation. 3. Evolution of the total number of effectively unpaired electrons of GO VII during the second-stage oxidation.

So dense covering of the basal plane by hydroxyls has resulted in a considerable distortion of the carbon skeleton structure that has been subordinated to the reconstruction of plane benzenoid-packed configuration into corrugated cyclohehanoid-packed one. A large family of cyclohexane isomorphs may explain a fanciful character of the skeleton structure. In contrast to the (5, 5) NGr hydrogenation [18], we were unable to distinguish particular kinds of cyclohexane isomorphs like arm-chair and boat-like that were observed at the hydrogenation of two-side accessible free standing molecular membrane. None can exclude that the cyclohexane isomorph definition is not suitable for the chemically modified graphene that offers a limitless number of possible accommodation to minimize energy losses.

Evolution of PCE in due course of the 52-step reaction at the second stage is shown in Fig.9. As seen, an oscillating behavior observed for PCE during the first stage of oxidation (Figs.5 and 7) is kept in this case as well. For comparison, the PCE plotting related to GO II is added. Comparing the data, one can see a drastic difference in the absolute PCE values in the two cases. Actually, the average values for GO II and GO VII constitute -102.15 and -44.79 kcal/mol, which indicates two-and-half-fold strengthening of hydroxyl coupling at their single addition to the molecule edge atoms in comparison with either the double additions to the edge atoms or single addition to atoms of the basal plane.

Plotting on the top of Fig.9 exhibits the evolution of the correlation of the (5, 5) NGr odd electrons in due course of the oxidation presented by changing in the molecular chemical susceptibility $N_D$. The plotting shows that as the number of the attached hydroxyls increases, the number of effectively unpaired electrons, of total number of 17.62 e for the pristine GO IV, gradually decreases, thus highlighting a gradual depletion of the molecular chemical ability. At the 52nd step $N_D$ becomes zero and keeps the value afterwards, which means stopping reaction at this step. The formed GO contains 74 hydroxyls, which provides 74:66, or 112.12 at% O:C ratio. It should be remained that H:C at% constitutes 130% (86:66) under the same conditions [18].

**Oxidation of carbonyl-framed GOs**

The first O-based SSHMR experiment, which involves the pristine GO I, can be attributed to homo-oxidant action with respect to both the first stage framing and further filling of the basal plane. As seen on the ACS image map in Fig.10a, the molecule framing by carbonyls suppresses the chemical activity of the molecule edge atom. Brightly shining balls in Fig. 1b are substituted by black ones with the only exclusion concerning atom 8 (circled) due more longer C=O bond in comparison with the others. The second stage oxidation starts at basal atom 43 (marked by oval in Fig.10a). The least total energy corresponds to the formation of an epoxy group that involves atom 43 and its neighbor. Checking this energy with respect to the position of oxygen atom either above or under the basal plane, we choose the best GO configuration following the lowest energy criterion. Selecting the next target atom by the ACS list of the GO, continue the procedure at the second step and so forth.

Presented in Fig.10b is the equilibrium structure of the 37-fold GO VIII that corresponds to the 15th step of two-side oxygen atom adsorption on the molecule basal plane. By this step, the initial pool of molecular chemical activity describing by the total number of effectively unpaired electrons ND equal 16.02 e for GO I is fully depleted, which stops the further reaction. The PSE evolution in due course of the stage is shown in Fig.11 (curve 1). For comparison, a similar plotting for the first stage framing by oxygen atoms is also added. The average PSE in these two cases constitute -51,31 and -135,65 kcal/mol, respectively. As in the previous case related to OH-GO VII, PSE decreases more than two times when the oxygen addition moves from edge atoms to the basal plane.

The experiment has been enlarged on hetero-oxidant adsorption on the basal plane. We have limited ourselves with O and OH since, as was shown, COOH is coupled very weakly. Figure 10c presents the 42-fold GO IX that corresponds to the 20th step of a successive attaching of either atomic oxygen or hydroxyl to the basal plane that is accessible to the oxidants from both sides. Criterion on the largest PSE has allowed for selecting the best configuration (O or OH as well 'up' or 'down'). The PSE evolution in due course of subsequent oxidation is given in Fig. 11. The average values constitute -38,7 and -48,3 kcal/mol for the OH and O additions, respectively. As seen in the figure, PSE curve 2 for the O addition is practically non-distinguishable from that obtained in the previous experiment. The PSE curve for the OH addition presents values less by absolute value than those for the O addition. However, in six cases the OH addition is more or less preferential so that the final GO IX contains six hydroxyls on the basal plane. The reaction is stopped at the 20th step due to depletion of the starting NB pool.

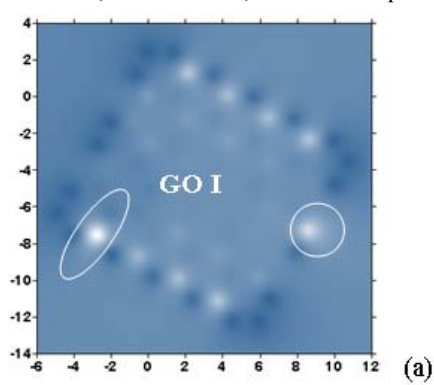

Oxidation of carbonyl-hydroxyl framed GOs

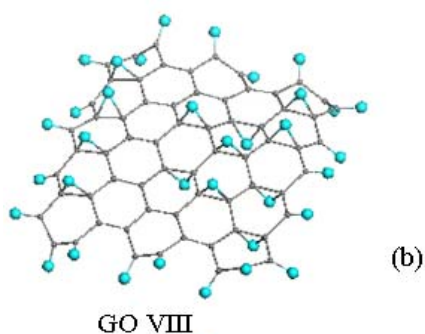

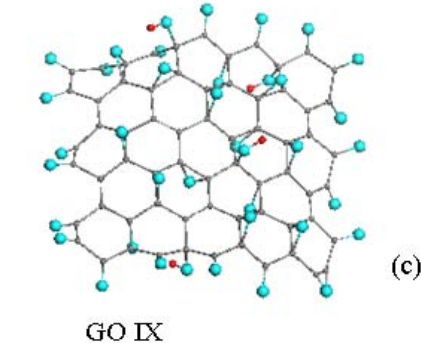

**Fig.10.** ACS image map of GO I (a) and equilibrium structures of 37-fold GO VIII (b) and 42-fold GO IX (c) obtained in due course of homo- and hetero-oxidant second-stage oxidations, respectively.

Described in previous sections has shown that atomic oxygen and hydroxyls are main rivals in regards the graphene molecule oxidation. Naturally to complete the consideration of GO topic by the description of what happens if both the first and second stages of the molecule oxidation involve the two oxidants. Filling the task, the last performed second-stage hetero-oxidant reaction (SSHTR) experiment concerns the oxidation of the (5, 5) NGr molecule that starts from the first-stage 22-fold GO V and continues by the addition of either atomic oxygen or hydroxyl to the basal plane of this oxide. Since under real conditions, one-side accessibility of the plane is the most favorable due to a top-down manner of graphite layer exfoliation [4, 6], we shall consider this case only.

Figure 12a presents starting conditions in terms of the GO V image ACS map. As seen in the map, the chemical activity of 20 edge atoms is fully compensated by the formation of carbonyl groups while two edge atoms at the upper and lower corners in the right are still active, comparable with the basal atoms, due to the attached hydroxyls saturate the atoms activity only partially. The reaction will start at basal atom 17 (circled). Attaching an oxygen atom causes the formation of an epoxy group that involves one of neighboring atom, besides atom 17. Attaching of hydroxyl terminates $N_{DA}$ of the atom. Similarly to the previous case, the selection of more profitable configuration at each step is governed by the largest PSE criterion. The oxidants addition to the molecule continues until the initial pool of effectively unpaired electrons $N_D$ equal 16,27 e is fully depleted.

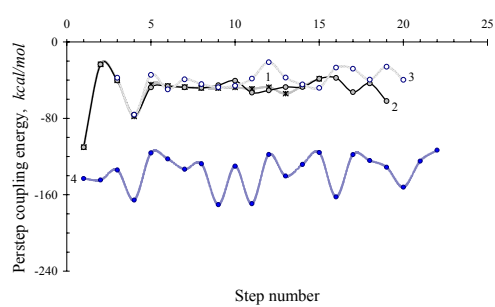

**Fig.11.** Perstep coupling energy versus step number for GO VIII family (1) and GO IX family (2. O-additios; 3-OH-additios) in due course of the second-stage oxidation. Curve 4 plots the same data for GO I family during the first-stage oxidation.

Equilibrium structure of the final configuration of the 43-fold GO X obtained at the 21st step of the second-stage subsequent oxidation is shown in Fig.12b. The final structure involves twenty carbonyls terminating the molecule edge atoms; four hydroxyls doing the same job, two of which are added during the second stage of the oxidation; fifteen epoxy and four hydroxyl groups randomly distributed over the basal plane. The total O:C at% ratio constitutes 65%. A particular attention should be given to a considerable curving of the carbon skeleton caused by the sp2 to sp3 transformation of the carbon valence electrons due to chemical saturation of their odd electrons. The structure transformation is similar to that one observed under one-side hydrogenation of (5, 5) NGr [18], albeit not so drastic.

The PSE evolution which accompanies the second-stage oxidation of GO V is presented in Fig.13. As seen in the figure, the O addition dominates and is characterized by an average PSE of -48,35 kcal/mol. Two zero points at curve 1 are related to the case when edge atoms previously attached by hydroxyls were the targets. The formation of epoxy groups assembling one edge and one basal atom was highly non-profitable while the addition of the second hydroxyl to the atoms compensated their free valence completely. Besides these two cases, the hydroxyl attachment was slightly favorable only in four cases more. The average PSE which is characteristic for hydroxyls constitutes -38,67 kcal/mol. For comparison, the average PSE for carbonyl-hydroxyl framing of GO V, presented by curve 3, is -133,76 kcal/mol.

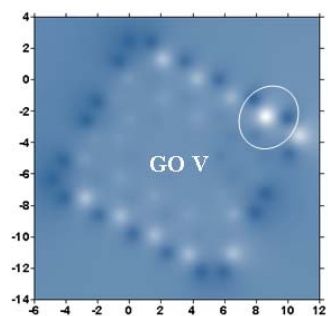

**Fig.12**. ACS image map of GO V (a) and equilibrium structures of 43-fold GO X obtained in due course of SSHTR oxidations.

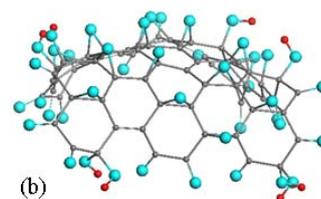

## Main GO characteristics and properties based on the computational experiment

A spatial configuration of the graphene molecule radicalization not only straightly points to a polyderivative character of any addition reaction with its participation but lays the foundation for the consideration of the polyderivatives formation computationally as a stepwise addition of the corresponding addend to the molecule carbon skeleton [49, 50]. Atomic chemical susceptibility (ACS), described as the number of effectively unpaired electrons on an atom NDA and determined at each step of the reaction, plays the role of a quantitative indicator of target atom, which is selected by the largest NDA value. This feature provides a possibility to trace any polyderivative formation sequentially. Applying the algorithm to the molecule oxidation and taking into account the presence of a number of oxidants, it is possible to trace the graphene molecule oxidation in two regimes, namely homo-oxidant and hetero-oxidant ones. The former implies the formation of homo-oxidant polyGOs in due course of all steps of the reaction dealing with each oxidant separately. When all the oxidants are considered one-by-one at each reaction step, hetero-oxidant polyGOs are created, the best GO of which is selected following the criterion of the largest PSE. The studied systems involve (5, 5) NGr molecule and three oxidants, namely atomic oxygen O, hydroxyl OH and carboxyl COOH. The obtained results show the following.

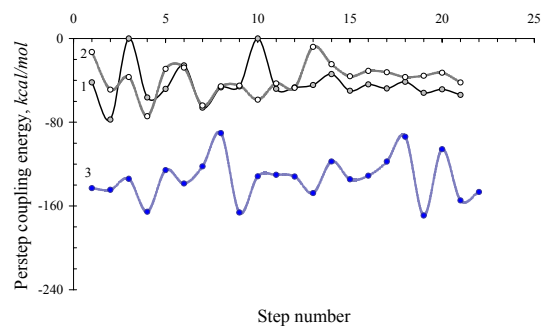

**Fig.13**. Perstep coupling energy versus step number for GO X family under O (curve 1) - and OH (curve 2) -additions in due course of the second-stage oxidation. Curve 3 plots the same data for GO V family during the first-stage hetero-oxidant reaction.

1. The graphene molecule is characterized by two zones of chemical reactivity which causes two-stage character of the oxidation. The first reaction zone covers edge atoms with high ACS values while the second zone involves atoms with much lower ACS that include all basal atoms and those edge atoms whose chemical reactivity was not suppressed in due course of the first-stage reactions. In view of the two-stage reaction, the oxidants are divided in two groups, the first of which concerns atomic oxygen while the second covers OH and COOH. The atomic oxygen fully suppresses the reactivity of edge atoms during the first stage of the reaction and deals with basal atoms only during the second stage. Two other oxidants leave the edge atoms with partially non-suppressed chemical reactivity during the first stage and concern both edge and basal atoms during the second stage.

2. The first-stage reaction concerns a 22-steps framing of the (5, 5) NGr molecule with the studied oxidants. Five fully framed GOs, namely, GOs I, II, III, IV, and V (see Figs. 2, 3, and 7) have been obtained in due course of the reaction, four first of which are homo-oxidant GOs, while GO V belongs to hetero-oxidant GOs. All the GOs correspond to one-point contact of oxidants with the molecule edge atoms thus providing the formation of C=O, C-OH, and C-COOH ending groups. In the case of the homo-oxidant O-framing, another possibility has been considered which provides the formation of epoxy ending groups involving one edge and one basal carbon atom each. Table 1 summarizes average PES and oxidant-induced chemical bonds data that characterize the obtained GOs.

**Table 1**. Average characteristics of the (5, 5) NGr molecule oxidation

| GOs[1] | PES, *kcal/mol* | | | C-O bonds lengths, $E$ | | | C-H bonds, $E$ |
|---|---|---|---|---|---|---|---|
| | O | OH | COOH | C-OH | O-C-O | C=O | |
| | | | | *The first-stage oxidation* | | | |
| I | -135,65 | - | - | - | - | 1,234 | - |
| II | - | -102,15[2] | - | 1,372 | - | 1,249[3] | 1, 101[3]; 0,971 |
| III | - | - | -77,34 | 1,361 | - | 1,233 | 0,972 |
| IV | - | -100.64 | - | 1,372 | - | - | 0,971 |
| V | -133.14 | -101,08 | -82,77 | 1,348 | - | 1,233 | 0,978 |
| Epoxy[4] | -0,70[5] | - | - | - | 1,369 | - | - |
| | | | | *The second-stage oxidation*[6] | | | |
| VI (II)[7] | - | -51,18[8] | - | 1,368 1,419 | - | 1,230[9] | 1, 103[9]; 0,970 |
| VII (IV) | - | -44.79 | - | 1,371 1,417 | - | - | 0,969 |
| VIII (I) | -51,31 | - | - | - | 1,434 | 1,234 | - |
| IX (I) | -50,44 | -40,25 | - | 1,417 | 1,438 | 1,223 | 0,970 |
| X (V) | -48,35 | -38,67 | - | 1,413 | 1,436 | 1,223 | 0,971 |
| XI (I) | -47,30 | -35,26 | -27,85 | 1,417; 1,354[10] | 1,437 | 1,230[10] | 0,971; 0,974[10] |

[1] Oxides under corresponding numbers are presented in Figs. 2, 3, 6,7, 9, and 11.
[2] The value is slightly overestimated due to one hydroxyl dissociation in due course of the reaction.
[3] Solitary C=O and C-H bonds, see Fig.2
[4] See the corresponding oxides in Fig.5
[5] The value is determined over first three step afterwards the PES becomes positive.
[6] The PES values are averaged over steps of the second stage only.
[7] Here and below figures in brackets point the reference to GOs of the first-stage reaction.
[8] The value has been determined over the first four steps of the second-stage oxidation due to termination of the reaction since the fifth stage (see Section 6.1).
[9] Solitary C=O and C-H bonds, see Fig. 7
[10] Data related to COOH oxidant.

As seen from the table, the average PES, which characterizes the formation of both carbonyl, carbo-hydroxyl and carbo-carboxyl ending groups, changes slightly when homo-oxidant addition is substituted by hetero-oxidant one. In all the cases, carbonyl ending is the most energetically preferable while carbon-carboxyl one occurs to be the least suitable. The last feature gives no chance for carboxyls to participate in the molecule framing when hetero-oxidant addition is possible. That is why there are no carboxyls among the oxidants framing of GO V. A drastic difference in the COOH heat of formation with respect to other oxidants discussed earlier undoubtedly lays the foundation of the feature. However, the COOH PES plottings in Figs. 5 and 7d have the largest amplitude and occur to be quite close to O- and OH-ones in spite of the lowest average PES while avoiding the intersection with the lowest one. This does not allow excluding COOH oxidant from the participation in framing the graphene molecule completely. It might occur that the PES plottings intersection can be influenced by the molecule size, which gives them a chance to be present. At any rate, their contribution will be small. On the other hand, more modest amplitude of the OH-plotting alongside with ~30% difference in the average PES with respect to carbonyl plotting favor the intersection of OH- and O-plottings and, thus, the presence of hydroxyls in a framing necklace. Evidently, the number of the intersection points depends on the molecule size due to which the OH contribution into the necklace is size-varied. Therewith, the average C-OH bonds lengths given in the Table 1 evidence a strong coupling of the oxidant at the molecule edges.

3. The second-stage oxidation concerns mainly the molecule basal atoms as well as those edge atoms whose initially high ACS was not fully suppressed by added oxidants during the reaction first stage. Six products of the reaction are presented by GOs VI, VII, VIII, IX, X, and XI. The average characteristics of GOs are listed in Table 1. The first two GOs are the products of the prolongation of the oxidation of GOs II and IV in due course of SSHMR. They can be regarded as general representatives of the OH-produced GOs. GO VIII presents an example of another SSHMR GO provided by atomic oxygen only. GOs IX, X, and XI are products of hetero-oxidant reaction concerning either the first reaction stage, or both and only one SSHTR, respectively. As seen from the table, the energy needed for each step of the second-stage oxidation decreases about three times for all the oxidants. This means that their coupling with the basal atoms is much weaker in comparison with interaction with edge atoms. The latter is supported by the elongation of all second-stage formed C-OH bonds and the formation of long O-C bonds of epoxy groups in comparison with rather short C=O bonds of ending carbonyls. A detailed description of chemical bonds in view of structural transformation of the GOs obtained is given in Supplementary Information, SI.

4. On the basis of the obtained results, is possible to suggest a reliable vision of the GO composition that meets the requirement of the top-down oxidation and is energetically profitable. Evidently, such GO, as all those discussed above, has two zones that differ compositionally. Thus, the framing zone predominantly consists of carbonyls accompanied by carbo-hydroxyls with much smaller contribution while the second basal zone is predominantly filled by epoxy groups with rather small contribution of hydroxyls. Such GO should look like GO X presented in Fig. 12. Computations have shown that carbo-carboxyl contacts are the least probable at both the first and the second stages of oxidation so that, if to be present they will constitute a definite minority in the GO final structures. As was already mentioned, the real shape and chemical composition of the GO will depend on the size of the pristine graphene molecule. Besides, the real oxidation processes are subordinated not only to static energy conditions but to kinetic requirements as well. The latter may change the contribution values of both carbonyl/epoxy component and hydroxyl one. However, since each individual addition of either O or OH to the molecule occurs barrierlessly [59], kinetics hardly changes the static results. As for COOH, the necessity to overcome a barrier at each individual addition to the graphene molecule [59] does not improve the situation with COOH discussed earlier.

5. Concerning GO morphology, it should be noted that the performed calculations have not revealed neither regularly structured GOs nor even those close to regularly structured ones. Moreover, the GO carbon skeleton has lost its planarity at early stage of the oxidation and becomes remarkably curved when the oxidation is terminated. Both features are due to $sp^2 \rightarrow sp^3$ transformation of the carbon electron system and cause the substitution of planar benzenoids of graphene by non-planar cyclohexanoids of GOs. A large variety of cyclohexanoid structures greatly complicated the formation of regularly structured graphene derivatives, in general, which is why the regular structure of graphene polyderivatives should be considered a very rare event. Until now it has been observed so far for a particularly arranged graphene hydride, one of a few, named graphane [17-19] while under the same conditions graphene fluorination does not produce derivatives with regular structure at all [18].

6. Fragmentation of the graphene polyderivatives is another important morphological feature. Actually, in the case of graphene oxidation, it was observed that the size of chemically produced GO is always less than that of the pristine graphite. Moreover, the GO size reduces when O:C at% ratio increases (see [85-87] and references therein). The feature is directly connected with a particular role of the edge zone for chemical interaction. If there is any topological reason for distinguishing a piece of the carbon atomic structure as an edge zone, strong chemical interaction with oxidant immediately fixes the area thus creating conditions for the zone enlarging due to anisotropy of the reconstructed odd electron cloud, on one hand, and mechanical stress, on the other.

## Some comments concerning GO reduction

The reduction of massively produced GOs is the second chemical reaction on the way of the transformation of dispersed graphite into a powder consisting, in an ideal case, of one-layer graphene flakes. Looking at GO X as a reliable model of GO in a general case and aiming at its returning to the graphene molecule shown in Fig.1, let us try to answer the following questions:

1. Is possible to return a drastically deformed and curved carbon skeleton structure of GO X to the planar structure of the (5, 5) NGr molecule?
2. Is possible to take out all the oxidant atoms from GO X making it free of dopants?

Computational experiment has an incomparable advantage of a free manipulation with atomic structure. Taking the opportunity we are able to release GO X from oxidant atoms just removing them from the output file and conserving the skeleton (core) structure therewith. Core structures X and XFR shown in Fig. 14a correspond to releasing GO X from all oxidant atoms in the first case and from oxidant atoms attached to the molecule during the second-stage of oxidation only, in the second. Skeleton structure VII in Fig. 14a demonstrates the core of GO VII after a similar releasing from all the OH groups. Afterwards, all the structures have been subjected to optimization. As seen in Fig. 14b, the optimization/reduction of cores X and VII fully restores the planar structure of the (5, 5) NGr molecule. The difference in the total energy of thus restored structures constitutes 0, 3% and 1.1% the energy of the (5, 5) NGr molecule relating to cores X and VII. Therefore, a drastic deformation of the GOs carbon skeletons is no obstacle for the restoration of the initial planar graphene pattern in spite of large deformational energy, namely, 874 kcal/mol and 1579 kcal/mol, incorporated in the deformed X and VII core structures. This is an exhibition of the extreme flexibility of the graphene structure that results in a sharp response by structure deformation on any external action, on one hand, and provides a complete restoration of the initial structure, on the other.

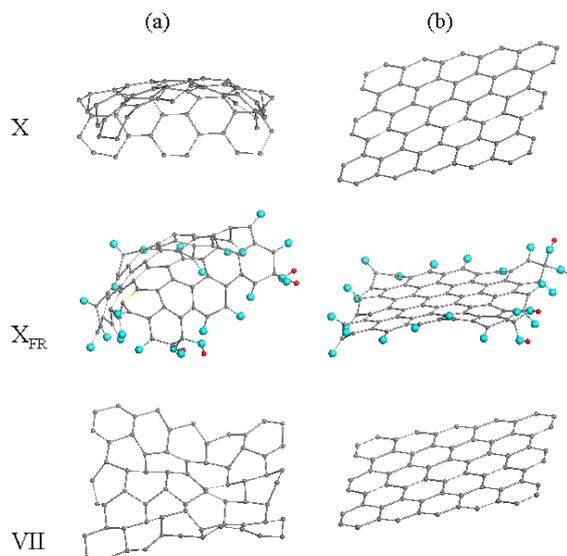

It should be noted, however, that not pure graphene but slightly oxidized product, called as rGO (see [4, 6, 58, 87] and references therein), is obtained after the GO reduction in practice. The equilibrium structure of core $X_{FR}$ in Fig. 14b may give a reliable presentation of the product. The difference in the total energy of the core and GO V, which should be considered as a pristine molecule in this case, constitutes ~15% which is much bigger than in the previous two cases. However, here, the matter is about the restoration of not a planar graphene structure but a non-planar flexible structure complicated by the presence of freely moving ending atoms. The obtained 15% losses of energy do not seem too high price for the transformation of the core $X_{FR}$ in Fig.14a into the structure that is practically non-distinguishable from that one of GO V.

**Fig.14**. Core structures of GO X and GO VII before (a) and after (b) structure optimization.

Chemists' activity in the GO reduction is insistently directed at the lowering of the dopant contribution in the rGOs. Oxygen has still remained the main impurity and is steadily recorded at amount not less than 5at% (see [4, 6] and references therein). We suppose that the analysis of the GO X structure and the data presented in Table 1 can explain this reality. Oxidant contribution into GO X is presented by carbonyl, carbo-hydroxyl and epoxy groups. The most strongly coupled are carbonyls located over the molecule perimeter. All other groups have about three times less PES and, consequently, are first candidates to be removed from the molecule. As known, carbonyls are highly stable up to high temperatures and can be removed as a whole just causing the destruction of the carbon skeleton. So that under conditions which conserve the skeleton structure, GOs can not be released from carbonyls on the flakes periphery just providing the rGO oxygen-contamination at the level roughly equal to the percentage of edge atoms. The latter constitutes ~1-2% for rGO flakes of ~0.1 in linear dimension and their contribution decreases when the molecule size increases further. However, each defect zone in the rGO body structure creates new 'edge' atoms that are terminated by atomic oxygen thus providing increasing oxygen contamination. As seen in Table 1, hydroxyls at the GO molecule edges are strongly bonded with the molecule core as well. Therefore, both carbonyls and ending hydroxyls may provide the main oxygen contamination of rGOs.

## Conclusive remarks and comparison with experimental data

Concluding the presentation of results of the described computational experiment, let us come back to the main hot points of the GO chemistry that have been still under question until now.

1. *Morphology*. Experiments reveal a remarkable disordering of the initial graphene structure even by partial oxidation so that chemically produced GOs are highly amorphous [4-16].

The performed computational experiment fully supports this finding and allows for explaining the feature by extremely high flexibility of the graphene structure and its highly sensitive response to any chemical modification.

2. *Graphene oxidation as a process in general*. Experimentally was shown that the oxidation of the graphene lattice proceeds in a rather random manner [4]. Graphene oxide is understood to be partially oxidized graphene [20]. The saturated at% ratio of oxygen to carbon is ~0.20-0.45 [7, 8, 21, 22]. When graphene oxide is heated to 11000 C, there is still about 5-10 at % oxygen left [15, 23-25].

All the features are supported by the performed calculations. As shown, the oxidation can be considered as a stepwise addition of oxidants to the pristine graphite/graphene body while the addition sequences for each monolayer are subordinated to a particular algorithm governed by the list of high-rank atomic chemical susceptibilities. In numerous cases presented in the current paper, was shown that the algorithm action does cause seemingly random distribution of oxidants over the pristine body in due course of the oxidation process. The algorithmic approach does not impose any restriction on the limit at% ratio of any addend, in general. This was supported by the results of the 'computational synthesis' of polyderivatives of fullerene $C_{60}$ [49] as well as polyhydrides and polyfluorides of graphene molecule (5, 5) NGr [18]. However, the performed computations, in full consent with previously considered hydrogenation of the molecule, have revealed that the initial radicalization of the molecule, which is provided by $N_D$ effectively unpaired electrons, has been gradually suppressed as the chemical reaction proceeds. The molecule chemical reactivity is gradually worked out approaching zero due to which the reactions stop. This explains why the hydrogenation and fluorination of fullerene $C_{60}$ is terminated at producing $C_{60}H_{36}$ and $C_{60}F_{48}$ polyderivatives, respectively, [52, 53] and why at% ratio of hydrogen to carbon in experiment of Elias et al. [17] is less than 133.3% when going from hydrides formed from two-side H-accessible perimeter fixed graphene membranes to one-side H-accessible graphene ripples [18]. The same regularities govern the graphene molecule oxidation, which, as shown, terminates the oxidation at achieving 112at% and 65at% of oxygen when the oxidation is provided by addition of either hydroxyls or oxygen atoms only. The latter situation is shown to be much more preferential. The saturation number involves filling of both edge and basal atoms. Since the model molecule is rather small, the contribution of edge atoms is significant. If exclude this contribution, the number of 48at% is characteristic for GO X that corresponds to the O-saturation of the basal plane mainly. The number is quite reasonable and points to a predominant $C_2O$ stoichiometry on the basal plane. Since experimental samples are much bigger in size, the mentioned earlier data of ~20-45at% are mainly related to the basal positions since the contribution of edge atoms is ≤1% at linear dimension of ≥0.1μ. In contrast, the availability of remaining oxygen in rGOs subjected to heating up to 11000C, is connected with edge atoms of rGOs. As shown, these atoms, which include not only perimeter atoms of the rGOs molecules but the atoms framing every defect zone, form a local area with very high chemical reactivity. Oxidants are strongly coupled with the atoms and can leave the molecule only alongside with carbon partners. The number of such atoms depends on linear size of both pristine GO molecules and their inner defects and may evidently constitute a few percents, which perfectly correlates with the observed amount of the remaining oxygen.

3. *Chemical composition of graphene oxide*. Basing on empirical data, the most common opinion attributes COOH, OH, and C=O groups to the edge of the GO sheet, while the basal plane is considered to be mostly covered with epoxide C-O-C and OH groups [4, 7, 9, 20].

The performed computations have allowed for forming up a hierarchy of the main three oxidants (O, OH, COOH) in regards their participation in the graphene oxidation that has shown an extremely low probability of such activity for carboxyls. Basing on the results obtained, it is possible to suggest a reasonable, self-consistent model of an ideal GO presented in Fig.15 Sure, the model cannot be simply scaled for adapting to larger samples. Obviously, due to extreme sensitivity of the graphene molecule structure and electronic system to even small perturbations caused by external factors, the fractional contribution of C=O, C-OH, and C-O-C groups may change in dependence of changing the molecule size, shape as well as of the presence of such impurities as metal atoms [88] and so forth. These facts may explain 'fluidness' of the term "graphene oxide" pointed by Ruoff et al [4].

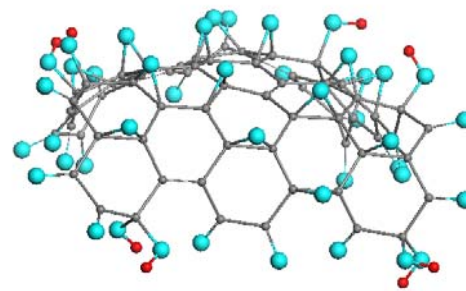

**Fig.15**. Structural model of a top-down exfoliated GO (43-fold GO X).

However, it is possible to convincingly state that the chemical composition of any GO has been governed by the presence of two zones drastically differing by the coupling of the relevant oxidants with the graphene molecule body so that dominant combinations of carbonyl/hydroxyl and epoxide/hydroxile will be typical for edge and basal areas of all GOs of different size and shape. Besides the chemical composition of chemically produced GOs, the performed calculations are able to suggest the chemical composition of rGOs as well. Discussion presented earlier and based on a two-zone-chemical-reactivity peculiarity of graphene molecules, clearly pointed to equilibrium $X_{FR}$ core in Fig. 14b as a reliable rGO model.

Concluding the discussion, we can state that the performed computational experiment has occurred to be able to consider practically all hot topics of the GO current chemistry and has offered explanation to all touched questions. Since the GO chemistry can be understood only through the whole entity of the topics, this explains why the performed experiment was so extended. The next step of computations has to be aimed at a detail consideration of the size-dependence of the obtained results and will obviously require a set of similarly large computational problems. Recently undertaken consideration of the first steps of oxidation of the (11, 11) NGr molecule, which exceeds (5, 5) NGr twice by linear size and four times by atoms number, has exhibited a similar behavior of both molecules.

Some words should be said concerning the computational strategy chosen in the study. Until now, the computational strategy of GO has been aimed at finding support to one of the available GO models, for which spectral data provide the main pool of data for comparison with calculations [35]. The strategy has been a result of certain limitations provided by a standard computational DFT scheme within the framework of solid-state periodic boundary conditions. However, the computational study of GO, based on such concept 'from a given structure to reliable properties' has resulted in the statement about kinetically constrained metastable nature of graphene oxide [35], thus revealing its inability to meet calls of the GO chemistry. In contrast, the molecular theory of graphene does not need any given structure beforehand but creates the structure in due course of calculations following the algorithms that take into account such fragile features of graphenes as their natural radicalization, correlation of their odd electrons, an extremely strong influence of structure on properties, a sharp response of the graphene molecule behavior on small action of external factors. Taking together, these peculiarities of the theory has allowed for getting a clear, transparent and understandable explanation of hot points of the GO chemistry discussed in the paper.

**Acknowledgements**

The authors wish to acknowledge the funding support for this project from the Russian Foundation for Fundamental Researches (grants 13-02-01-076 and 13-08-00639).

# Molecular theory of graphene oxide

**Elena F. Sheka,\* Nadezhda A. Popova**

*Peoples Friendship University of Russia*

117198 Moscow, Miklukho-Maklay 6, Russia

\*To whom correspondence should be addressed. E-mail: sheka@icp.ac.ru

Oxidation-induced structure transformation. Stepwise oxidation is followed by the gradual substitution of $sp^2$-configured carbon atoms by $sp^3$ ones. Since both valence angles between the corresponding C-C bonds and the bond lengths are noticeably different in the two cases, the structure of the carbon skeleton of the pristine (5,5) NGr molecule loses its planarity and becomes pronouncedly distorted.

Figs.S1a and b demonstrate the transformation of the skeleton structure in due course of both first-stage and second-stage oxidation exemplified by changes within a fixed set of C-C bonds of GO IV and GO VII. Comparing the pristine diagram with those belonging to a current oxidized species makes it possible to trace changes of the molecule skeleton structure. As seen in Fig. S1a, the OH-framing of GO IV causes a definite regularization of the C-C bonds distribution. None of single C-C bonds is formed due to incomplete suppression of the chemical activity of edge atoms due to which $sp^2 \rightarrow sp^3$ transformation does not occur. Contrary to this, the second-stage oxidation is followed by the formation of single C-C bonds that evidently prevail in Fig.S1b thus exhibiting a large-scale $sp^2 \rightarrow sp^3$ transformation. The appearance of elongated C-C bonds, the number of which increases when oxidation proceeds, is naturally expected. However, to keep the skeleton structure closed, this effect as well as changes in valence angles should be compensated. At the level of bonds this compensation causes squeezing a part of pristine bonds. Actually, the plotting in Fig.S1b shows seven C-C bonds which are very short and whose lengths are in the interval of 1.345-1.355 Å. These lengths point to non-saturated valence state of the relevant carbon atoms which are not attached by oxygen, on one hand, however, on the other hand, are under the critical value of 1.395Å that put a lower limit for initiation of the odd electron correlation [1]. Under this value, the bond length provides a complete covalent bonding of odd electrons due to which these electrons are completely correlated so that the total number of effectively unpaired electrons $N_D$ is zero. The formation of these short bonds explains stopping of the oxidation reaction at the $52^{th}$ step of the second-stage oxidation.

Figure S1c presents the length distribution of C-OH and O-H oxidant given groups. Dark red plotting is related to GO IV and describes the bonds of framing atoms, while gray plotting shows similar distribution in the case of the 74-fold GO VII. As seen in the figure, the bonds are quite regularly distributed in the both cases. The relevant dispersion of the bonds is listed in Table S1. The dispersion is bigger in the latter case but does not exceed 2% of the average value pointed in the table. As seen in the figure, all the C-OH bonds related to hydroxyl attached to the basal plane atoms are longer than those related to OH-framing of the molecule during the first stage of oxidation (1.417 and 1.371Å, respectively, see Table S1). Important to note that previously short C-OH bonds related to edge atoms elongate during the second stage of oxidation as well and approach the average length of 1.417 Å. At the same time, part of these bonds, which remain untouched during this stage, preserve their shorter lengths.

The transformation of the molecule skeleton structure in the case of hetero-oxidant reaction is shown in Fig.S2. Figure S2a presents the skeleton distortion of GO V by the end of the first-stage oxidation. Since 20 carbonyls and two C-OH groups form the molecule framing, the dominant elongation of C-C bonds due to $sp^2 \rightarrow sp^3$ transformation provided by carbonyls is clearly seen Two C-OH groups leave the corresponding edge atoms in sp2 configuration. The second-stage oxidation concerns the two edge and 38 basal atoms.

Shown in Fig. S2b exhibits the skeleton distortion by the 21st step of the second-stage oxidation for GO X. The dominant majority of bonds become single due to large-scale sp3 sp2 transformation. The remaining untouched carbon atoms form bonds of 1.343-1.358Å in length similarly to the case shown in Fig.S1b. Reasons for these bonds creation are discussed above. As previously, the bonds length provides a complete covalent bonding of the remaining odd electrons thus conserving sp2 configuration of the relevant carbon atoms and terminating a further oxidation. The situation related to GO VII and GO X is quite analogous to those that take place in due course of the stepwise hydrogenation [2] and fluorination [3] of fullerene C60.

Figure S2c discloses events that concern attached oxidants. As said above, the first-stage oxidation is terminated by a complete framing of the pristine molecule by 20 carbonyls and two hydroxyls. Consequently, dark red curve in Fig.S2c exhibits 20 C=O bonds as well as two C-OH and two O-H bonds of GO V. Gray curve presents the bond structure after terminating of the second-stage oxidation for GO X at the 21st step. As seen, previously much shorter C-OH bonds of 1.348Å lengthen up to 1.413Å and are comparable by length with all the newly formed C-OH bonds; C=O bonds differ slightly, and C-O-C bonds of epoxide groups are added. The average length values are listed in Table S1 alongside with the bond length dispersion. As seen in the table, the bond distributions are quite homogeneous and their dispersion does not exceed 1%.

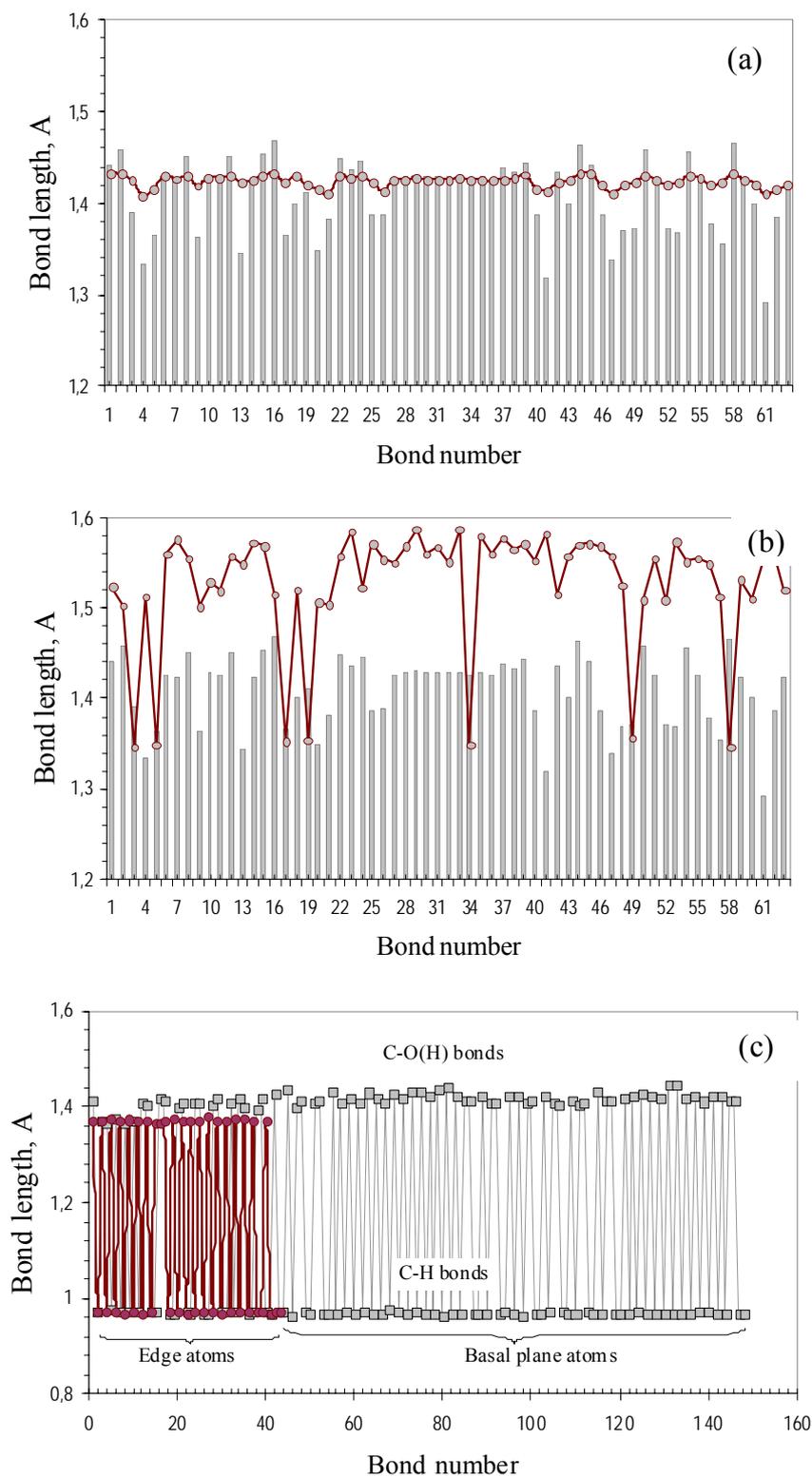

**Fig.S1**. sp$^2$→sp$^3$ Transformation of the (5, 5) NGr molecule skeleton structure in due course of successive oxidation. a. First-stage oxidation, GO IV; b. Second-stage oxidation, GO VII. Light gray histogram is related to the pristine molecule. c. Oxidant-generated groups at the first-stage (dark red) and the second-stage (gray) oxidation.

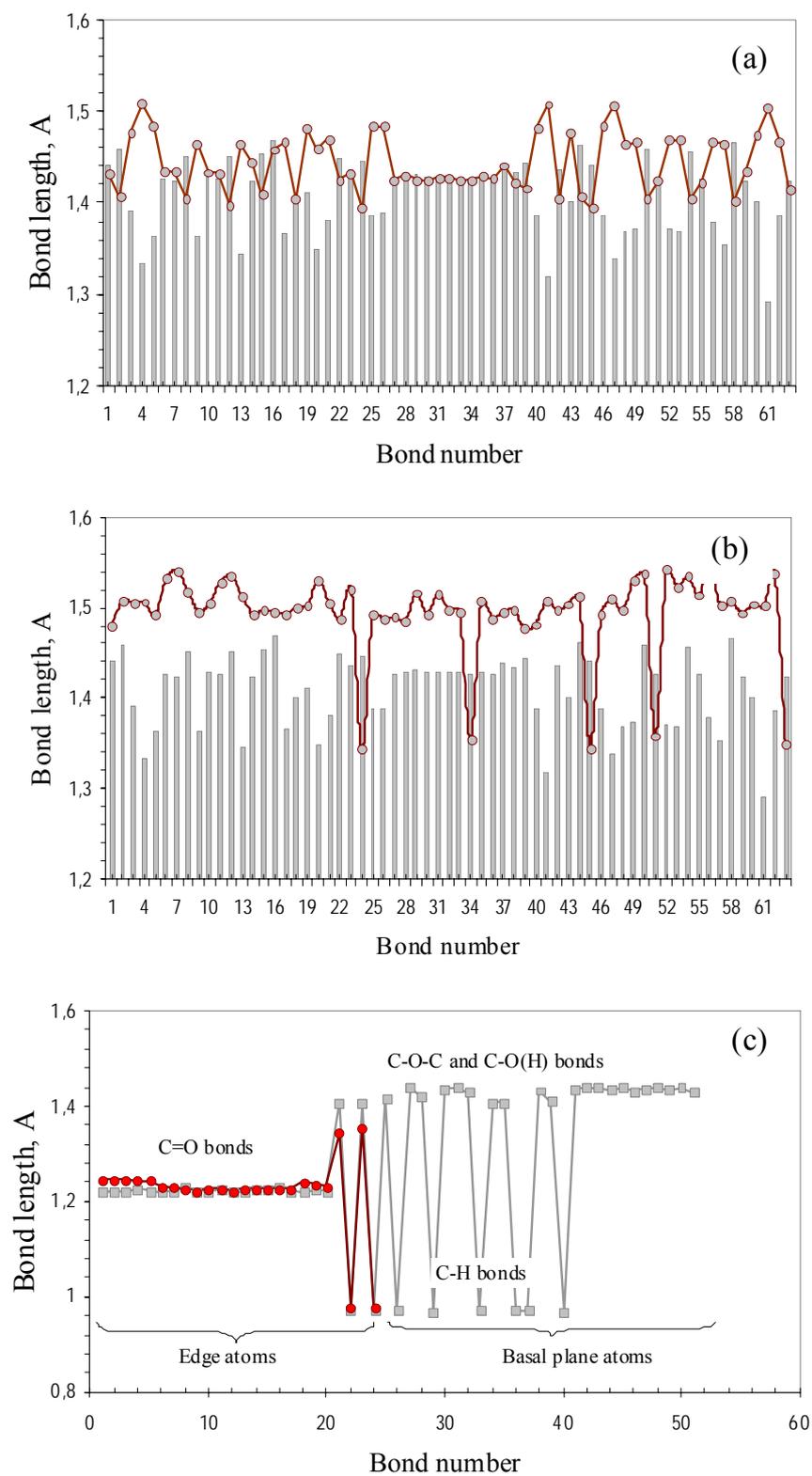

**Fig. S2.** sp$^2$→sp$^3$ Transformation of the (5, 5) NGr molecule skeleton structure in due course of successive oxidation. a. First-stage oxidation, GO V; b. Second-stage oxidation, GO X. Light gray histogram is related to the pristine molecule. c. Oxidant-generated groups at the first-stage (red) and the second-stage (gray) oxidation.

**Table 1S**. Average bond lengths of the (5, 5) GOs and their dispersion, %

| GOs[1] | C-O bonds lengths, $E$ | | |
|---|---|---|---|
| | C-OH | O-C-O | C=O |
| First-stage oxidation | | | |
| I | - | - | 1,234 (5.30; - 0,90) |
| II | 1,372 (0.70; -0,30) | - | 1,249[2] |
| III | 1,362 (3.79; -0.99) | - | 1,233 (0.58; -0.47) |
| IV | 1,372 (0.46; -0,33) | - | - |
| V | 1,348 (0.43; -0.43) | - | 1,233 (1.04; -0.83) |
| Second-stage oxidation | | | |
| VI (II)[3] | 1,368 (2.10; -4.01); 1,419 (1.85; -1.39) | - | 1,230[4] |
| VII (IV) | 1,371 (0.32; 0.18) 1,417 (1.98; -1.68) | - | - |
| VIII (I) | - | 1,434 (0.20; -0.27) | 1,234 (0.27; -0.21) |
| IX (I) | 1,417 (0.29; -0.14) | 1,438 (0.38; -0.46) | 1,223 (0.42; -0.38) |
| X (V) | 1,413 (1.08; -0.46) | 1,436 (0.24; -0.46) | 1,223 (0.63;-0.30) |
| XI (I) | 1,417; 1,354[5] (0.24;-0.22) | 1,437 | 1,230[5] (0.25; -0.27) |

[1] Oxides under corresponding numbers are presented in Figs. 2, 3, 7,8, 10, and 12.
[2] A solitary C=O bond, see Fig.2
[3] Here and below figures in brackets point the reference to GOs of the first-stage reaction.
[4] A solitary C=O bond, see Fig. 8
[5] Data related to COOH oxidant.